# High-order alloharmonics produced by nonperiodic drivers

M. S. Pirozhkova[1*], K. Ogura[1*], A. Sagisaka[1], T. Zh. Esirkepov[1], A. Ya. Faenov[2†], T. A. Pikuz[2],
H. Kotaki[1], Y. Hayashi[1], Y. Fukuda[1], J. K. Koga[1,7], S. V. Bulanov[1,3], H. Daido[4], N. Hasegawa[1],
M. Ishino[1], M. Nishikino[1], M. Koike[1], T. Kawachi[1], H. Kiriyama[1], M. Kando[1], D. Neely[5,6†],
A. S. Pirozhkov[1‡]

[1] *Kansai Institute for Photon Science (KPSI), National Institutes for Quantum Science and Technology (QST), 8-1-7 Umemidai, Kizugawa, Kyoto 619-0215, Japan*
[2] *Institute for Open and Transdisciplinary Research Initiatives, Osaka University, Suita, Osaka 565-0871, Japan*
[3] *Institute of Physics of the Czech Academy of Sciences, Extreme Light Infrastructure Beamlines Project, Na Slovance 2, 18221 Prague, Czech Republic*
[4] *Institute for Laser Technology, 2-6 Yamada-oka, Suita, Osaka 565-0871, Japan*
[5] *Central Laser Facility, Rutherford Appleton Laboratory, STFC, Chilton, Didcot, Oxon OX11 0QX, United Kingdom*
[6] *Department of Physics, SUPA, University of Strathclyde, Glasgow G4 0NG, United Kingdom*
[7] *Kyoto International University Academy, 63-1 Yuden Tanabe, Kyotanabe Kyoto 610-0331, Japan*

High-order harmonics are ubiquitous in nature and present in electromagnetic, acoustic, and gravitational waves. They are generated by periodic nonlinear processes or periodic high-frequency pulses. However, this periodicity is often inexact, such as that in chirped (frequency-swept) optical waveforms or interactions with nonstationary matter – for instance, reflection from accelerating mirrors. Spectra observed in such cases often contain complicated sets of harmonic-like fringes, uninterpretable or even misinterpretable via standard Fourier analysis. Here, we propose the concept of *alloharmonics*, i.e. spectral interference of harmonics with different orders, fully explaining the formation of these fringes (from Greek ἄλλος: állos, "other"). Like atomic spectra, the complex alloharmonic spectra depend on several integer numbers and bear a unique imprint of the emission process, such as the driver period and its time derivatives, which the alloharmonic theory can decipher. We demonstrate laser-driven alloharmonics experimentally in the extreme ultraviolet spectral region and extract nonperiodicity parameters. We analyze previously published simulations of gravitational waves emitted by binary black hole mergers and demonstrate alloharmonics there. Further, we predict the presence of alloharmonics in the radio spectra of pulsars and in optical frequency combs, and propose their use for measurement of extremely small accelerations necessary for testing gravity theories. The alloharmonics phenomenon generalizes classical harmonics and is critical in attosecond physics, frequency comb generation, pulsar studies, and future gravitational wave spectroscopy.

---

[*] These authors contributed equally to this work
[†] Deceased
[‡] Contact author: pirozhkov.alexander @ qst.go.jp

# I. INTRODUCTION

## A. High-order harmonics

High-order harmonics are coherent spectra with distinct components at angular frequencies $\omega_n=(n+\delta n)\omega_0$, where $\omega_0$ is the driver (also called base or fundamental) frequency, the integer $n$ is the harmonic order, and $0 \leq \delta n < 1$ corresponds to the carrier-envelope offset. Harmonics appear in many processes, such as laser-driven harmonic generation in ionizing gases [1] and from plasma surfaces [2] and burst intensification by singularity emitting radiation (BISER) [3, 4]. Similar spectra appear in the most accurate absolute frequency references [5, 6] – optical frequency combs, i.e., spectra of highly periodic optical pulse trains generated by a laser pulse bouncing inside stabilized cavities. Harmonics are also present in acoustic and gravitational waves [7, 8]. Some orders may disappear due to specific (e.g. field reversal) symmetries: gas harmonics generated by single-frequency multi-cycle drivers contain odd orders (harmonics separation $\Delta\omega=2\omega_0$), while plasma surface harmonics may contain all ($\Delta\omega=\omega_0$), odd or even (both $\Delta\omega=2\omega_0$) orders depending on the incidence angle and polarization [9, 10]. Importantly, harmonic separations are always integer *multiples* of the driver frequency.

Equivalently, harmonics can be considered in the time domain connected with the spectral domain via the Fourier transform. Periodically spaced spectral components with separation (period) $\Delta\omega$ correspond to coherent periodically spaced pulses in time separated by $\Delta t=2\pi/\Delta\omega$. For harmonics spaced by $k$ driver frequencies, the pulse spacing is $\Delta t=2\pi/k\omega_0=T_0/k$, where $T_0$ is the driver period. In a frequency comb, the time separation $\Delta t=2\pi/\omega_0=1/\nu_{\text{rep}}$ is the inverse of the pulse train repetition rate. Harmonics are often nearly-phase-matched, and their Fourier synthesis gives sharp pulses in time with durations considerably shorter than the driver period. This generates femtosecond laser pulses – in research, technology, and medicine – employing $\sim 10^5$ to $10^7$ harmonics with MHz or GHz frequency combs, and attosecond pulses – the fastest artificial probes [11] – employing several to several tens of harmonics with various generation mechanisms [12-15].

## B. Nonperiodic drivers

The harmonic-generating driver frequency is – or assumed to be – constant. Yet, an exact driver periodicity is rather an exception. There are many important inexactly periodic drivers. For instance, a frequency sweep (chirp) appears due to a time-varying Doppler shift upon nonlinear reflection from accelerating mirrors [16-19]. Chirped pulse amplification [20] allows generating the highest laser powers [21]. Binary black holes' rotation gradually accelerates during inspiral [8, 22]. Pulsars' rotation slowly decelerates [23, 24] but accelerates during glitches [25]. The repetition rates of the pulse trains forming frequency combs [5, 6] would change if the laser cavity parameters drift or are intentionally varied. It is known that high-frequency spectra generated by such nonperiodic drivers differ in various ways from classical harmonics. For instance, the bandwidth of each individual harmonic can increase with order [26], or half-integer harmonics can appear [18], or harmonics can disappear, being replaced with harmonic-like fringes [19]. In each case, it has been correctly reasoned that the observations are caused by driver nonperiodicity.

However, like atomic spectra in the pre-quantum-mechanics-era, this variety of observations in its entirety cannot be explained by current concepts and models, as we show via our experimental example (Section II.A). Here we propose a concept and theory of coherent spectra formation in nonperiodic cases, which we call *alloharmonics*. We explain our experimental data and their former inexplicability. The theory reproduces previously known cases and allows predicting and explaining all the puzzling features like spectral fringe periods approximately equal to various integer *fractions* of the driver frequency. The concept is widely applicable to all nonperiodic harmonic-generating drivers: lasers, frequency combs, radio-pulsars, and gravitational waves from binary mergers; in all cases previously unavailable nonperiodicity properties can be extracted from alloharmonic observations. As discussed later alloharmonic theory could also be the basis for the scaling necessary for laboratory astrophysics [27, 28]. On the other hand, unaccounted-for alloharmonics can cause misinterpretation of observations in laser labs and astrophysics.

## II. RESULTS

### A. Experiment

In the experiment an intense laser pulse (irradiance $\sim 10^{19}$ W/cm$^2$) propagating in tenuous plasma (electron density $\sim 3 \times 10^{19}$ cm$^{-3}$) produced structurally stable relativistic singularities (Appendix A, Sections A-E). These nanoscale-size sources emitted bright coherent extreme ultraviolet radiation via BISER [3, 4], FIG. 1(a). The raw data and spectrograph calibration and resolving power are shown in Appendix A, Section F. The spectrum exhibits various fringes, and in narrow spectral ranges would appear as some harmonics with different spectral fringe spacing and carrier-envelope offset: $\omega_n \approx n \times 0.63$ eV, $\omega_n \approx n \times 0.3$ eV, or $\omega_n \approx (n+0.5) \times 0.63$ eV, depending on the region, FIG. 1(b-g), with all fringe periods different from the laser photon energy, $\hbar\omega_L = 1.53$ eV. The entire spectrum could not be interpreted in terms of harmonics. However, it is known that the intense laser frequency changes over time due to propagation in an ionizing medium [29, 30] and photon acceleration [31-33]. Also, the laser frequency continuously and gradually downshifts in the nonlinear depletion process [34-36], where the laser pulse transfers its energy to plasma waves, which in turn power laser electron accelerators [37], BISER, and other bright x-ray sources [38], and produce intense single-cycle infrared pulses [39, 40]. Thus, we inferred that this spectrum was generated by a nonperiodic driver. However, the observed spectral features remained unexplained and nonperiodicity parameters remained unknown. As in astrophysical observations, a complication was that at the instant of BISER generation the laser frequency, which in other laser experiments is a reliable reference, was unknown.

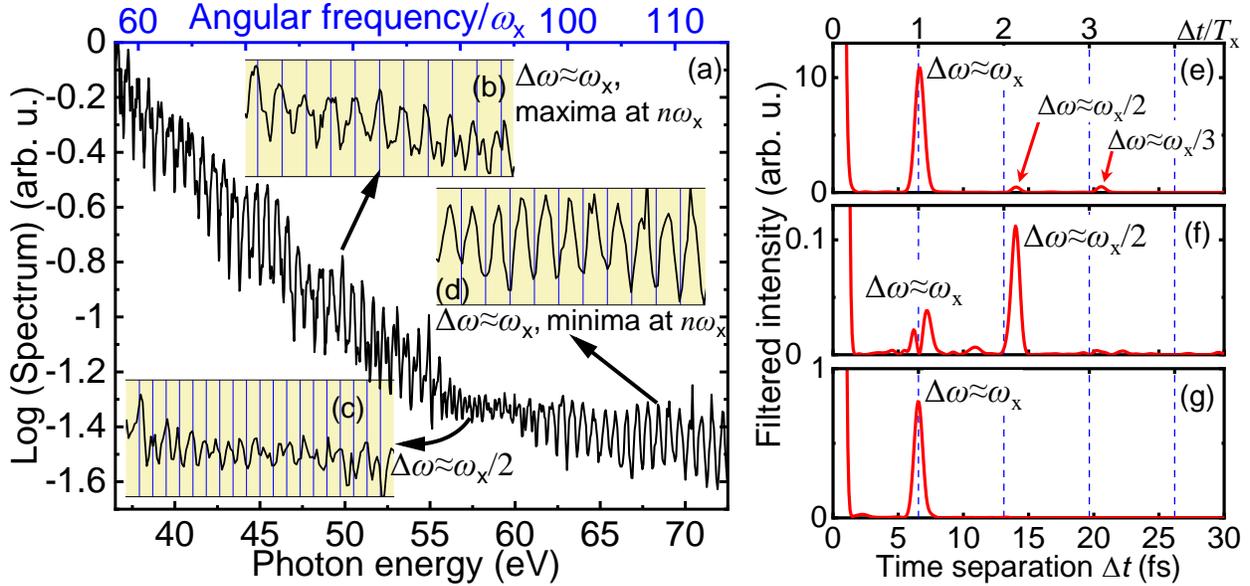

FIG. 1. Coherent extreme ultraviolet spectrum generated by a gradually-frequency-downshifting relativistic-irradiance laser pulse via the burst intensification by singularity emitting radiation (BISER) mechanism. (a) Experimental BISER spectrum. (b-d) Spectral regions with fringe separations $\Delta\omega \approx \omega_x$, $\omega_x/2$, and $\omega_x$, respectively, with $\hbar\omega_x \approx 0.63$ eV; vertical lines indicate integer multiples of the corresponding fringe periods, $n\Delta\omega$. Note that the spectral fringe period in (c) differs from that in (b) and (d), and all the fringe periods differ from the initial laser photon energy $\hbar\omega_L \approx 1.53$ eV. Also note that at frequencies $n\omega_x$ the spectrum exhibits antinodes in (b) but nodes in (d). (e-f) Intensities of local Fourier transforms of the experimental spectrum with centers at $\hbar\omega$=45, 58, and 66 eV, corresponding to (b-d), respectively (Gaussian spectral windows with 2 eV widths). Blue vertical dashed lines denote periods corresponding to $\omega_x$.

## B. Alloharmonics

Figures 1(e-g) show local Fourier transforms: the discrete peaks demonstrate time separations that determine local fringe periodicity. Like harmonics, the experimental fringes resulted from spectral interference of coherent time-separated pulses in a train, but, in contrast to harmonics, the pulse separations, or, in other terms, the driver frequency, were nonconstant. However, the spectral interference requires equivalence (overlap) of frequencies of different pulses, which seems impossible due to the frequency change: $(n+\delta n)\omega \neq (n+\delta n)\omega'$ for $\omega \neq \omega'$. To explain the experimental observations, we propose the *alloharmonic concept*, i.e., the spectral interference of the harmonic order $n$ from pulse number $m$ of the pulse train (instantaneous frequency $\omega_m$) with *another* harmonic $n+\Delta n$ from pulse $m+\Delta m$ (instantaneous frequency $\omega_{m+\Delta m}$), with the interharmonic separation $\Delta n$ and pulse separation $\Delta m$>0. The frequency equivalence gives *the alloharmonic equation*:

$$(1) \qquad (n+\delta n)\omega_m = (n+\delta n+\Delta n)\omega_{m+\Delta m}.$$

This is a linear equation for $n$ yielding a nontrivial solution for $\omega_m \neq \omega_{m+\Delta m}$. The alloharmonic fringes appear at corresponding frequencies $\omega_A = (n+\delta n)\omega_m$.

In many important cases the dependence of the driver frequency on time $t$ can be written or approximated as

$$(2) \qquad \omega(t) = \omega_0 \left(1 + 2\alpha \frac{t}{T_0} + 3\beta \left(\frac{t}{T_0}\right)^2 + 4\gamma \left(\frac{t}{T_0}\right)^3 + 5\delta \left(\frac{t}{T_0}\right)^4 + \cdots \right)$$

with dimensionless nonperiodicity (chirp) parameters $\alpha, \beta, \gamma, \delta, \ldots$ In the small nonperiodicity case, these chirp coefficients are small. A detailed model description and equation derivations are provided in Appendix B and Supplemental Material [41].

For example, in the linear chirp case,

$$(3) \qquad \omega(t) = \omega_0 \left(1 + 2\alpha \frac{t}{T_0}\right),$$

FIG. 2, the nonperiodic pulse train has individual pulses at $t_m = \frac{\sqrt{1+4m\alpha}-1}{2\alpha}T_0 \approx (m - \alpha m^2)T_0$ with frequencies $\omega_m = \omega(t_m) = \omega_0\sqrt{1+4m\alpha} \approx (1+2m\alpha)\omega_0$, FIG. 2(a); here and below the approximations are valid for the small nonperiodicity case. The overall spectrum is complicated, FIG. 2(b). At relatively low frequencies, <25$\omega_0$ in this example, there are harmonics at frequencies $(n+\delta n)\omega_0$, FIG. 2(c), due to spectral interference of the same-order harmonics of different pulses, thus forming the $\Delta n$=0 series, the *approximate* harmonic solution of Eq.(1). These harmonics differ from the ideal (periodic) harmonics as their amplitudes and visibility gradually decrease and widths increase with order. However, starting from some frequency $\omega_{min}$, there appear additional fringes: These are the *alloharmonics*, FIG. 2(d). At some higher frequency $\omega^*$, the harmonics disappear and alloharmonics become the dominant spectral feature, FIG. 2(e,f). For the linear chirp case and Gaussian temporal envelope with duration $\tau$, these frequencies and the corresponding orders are [41]

$$(4) \qquad n_{min} \approx \frac{\omega_{min}}{\omega_0} \approx \left|\frac{T_0}{4\alpha\tau}\right| = \left|\frac{\pi}{2\alpha\omega_0\tau}\right| \text{ (alloharmonics appear)},$$

$$n^* \approx \frac{\omega^*}{\omega_0} \approx \left|\frac{T_0}{2\alpha\tau}\right| = \left|\frac{\pi}{\alpha\omega_0\tau}\right| \text{ (alloharmonics dominate)}.$$

These equations can also be used as estimates for non-Gaussian envelopes and higher-order chirps with a dominant linear term. As $|\alpha|$ increases, the alloharmonics shift towards lower frequencies, FIG. 2(g) and the Supplemental Movie [42].

Alloharmonics appear at sets of frequencies which satisfy Eq.(1) for different integer values of $m$, $\Delta m > 0$, and $\Delta n \neq 0$. Eq.(1) can be illustrated by a moiré pattern, FIG. 2(h): like a vernier scale, the two different combs match at certain different line numbers (harmonic orders), thus allowing spectral interference. Such moiré patterns differ for each $(m, \Delta m)$ pair; thus, alloharmonics appear at many different frequencies $\omega_A$ and form much more complicated fringes than suggested by a single moiré pattern. For the linear chirp (3), Eq.(1) yields:

$$(5) \quad \frac{\omega_A}{\omega_0} \approx -\frac{1}{2\alpha}\frac{\Delta n}{\Delta m} - 3\Delta n \left(\frac{m}{\Delta m} + \frac{1}{2}\right).$$

The first term dominates for $\alpha \ll 1$, resulting in several closely-spaced series with equal ratios $\Delta n/\Delta m$ [dashed vertical lines in FIG. 2(b-f)]. We note that $\Delta n$ and $\alpha$ have opposite signs and $\Delta m > 0$, so the first term in (5) is always positive.

The alloharmonic fringe periods for the linear chirp (3) are $\Delta \omega \approx \frac{\omega_0}{\Delta m} + \left(1 + \frac{2m}{\Delta m}\right)\alpha \omega_0$. For $\alpha \ll 1$, these are $\Delta \omega \approx \omega_0/\Delta m$, consistent with our experiment, FIG. 1, and with FIG. 2(d-f), exhibiting $\Delta \omega \approx \omega_0/6$, $\omega_0/3$ and $\omega_0/1$ fringe periods, respectively. We note that the approximate equality $\Delta \omega \approx \frac{\omega_0}{\Delta m}$ holds not only for the linear chirp case, but in general for all cases of small nonperiodicity, Appendix B Section B.

There exist alloharmonics with $\Delta m = 1$ and fringe period *not exactly* $\omega_0$ and spectral peaks *not* at the harmonic positions $(n+\delta n)\omega_0$. In the linear chirp case, they appear at frequencies

$$(6) \quad \omega_{A1} \approx -\frac{\Delta n}{2\alpha}\omega_0$$

[FIG. 2(f) at $\omega_{A1} \approx 165\omega_0$ and dotted lines in FIG. 2(g)].

The alloharmonics can be equivalently considered from another perspective. Due to the finite envelope duration and nonperiodicity, the spectrum of a single harmonic has a finite, increasing with the harmonic order, bandwidth. The higher-frequency part of the spectrum is considerably more sensitive to nonperiodicity, FIG. 2(g) (see also FIG. 7(c) in Appendix B and the Supplemental Movie [42]): the low-order harmonics are narrow even at $\alpha = -0.01$, while higher-order harmonics are much wider and start to overlap: *this produces the interharmonic interference, i.e., alloharmonics*. Thus, solutions of the alloharmonic equation, Eq.(1), can be non-integer so the fringes appear between the ideal harmonics. For a linear chirp and Gaussian envelope, the bandwidth of a single harmonic is $\delta \omega_n = \frac{2}{\tau}\sqrt{1 + 4\pi^2\alpha^2(n+\delta n)^2 \frac{\tau^4}{T_0^4}}$ [41]. At low orders, there are normal narrow-bandwidth nonoverlapping harmonics. At order $n_{\min}$ [Eq.(4)], the widths of the individual harmonics become $\delta \omega_{n,\min} \approx 0.5\omega_0$, and neighboring harmonics start to overlap, producing alloharmonic fringes in the minima between the harmonics. At order $n^*$ [Eq.(4)], $\delta \omega_{n^*} \approx \omega_0$, neighboring harmonics overlap completely and disappear, while alloharmonics dominate. Details may differ for other envelope shapes, e.g., for a flat-top temporal envelope, alloharmonics start to appear on top of the harmonic peaks.

Importantly, the chirp rate $\alpha$ and overall duration $\tau$ occur as products in the denominators in Eq.(4). Even for slightly inexact periodicity, there are still frequencies at which the alloharmonics become dominant. In our example, the relative frequency change during the entire pulse $\sim 2\alpha\tau/T_0$ is less than 2%, but the alloharmonics became dominant at $\omega^* \sim 56\omega_0$. In frequency comb spectroscopy and pulsar studies, the envelope duration $\tau$ equals observation time, which can be hours. Thus, alloharmonics appear even with extremely slow frequency changes. Further, for $\tau \gg T_0$, there are many pulses in the train, so the pulse separation $\Delta m$ can be very large, leading to very fine fringes; their observation requires high resolving power $R=\omega_A/\Delta\omega$, which is $\sim\Delta m$ times higher than for harmonics [41]. For the linear chirp,

$$(7) \quad R = -\frac{\Delta n\sqrt{1-16\alpha^2(\tau/T_0)^2}}{2\alpha} \approx -\frac{\Delta n}{2\alpha}.$$

In the lowest-order approximation, the required resolving power does not depend on the selected spectral range because the spectral fringe period in this approximation depends linearly on the frequency as $\Delta\omega \approx -\frac{2\alpha}{\Delta n}\omega$ [41]. In our example, in the spectral region around $\omega=28\omega_0$, FIG. 2(d), the minimum resolving power for observation of still-visible harmonics is $R_{HH}=28$, while for the starting-to-appear $\Delta n=1$ alloharmonics is $R_{min}\approx 167$.

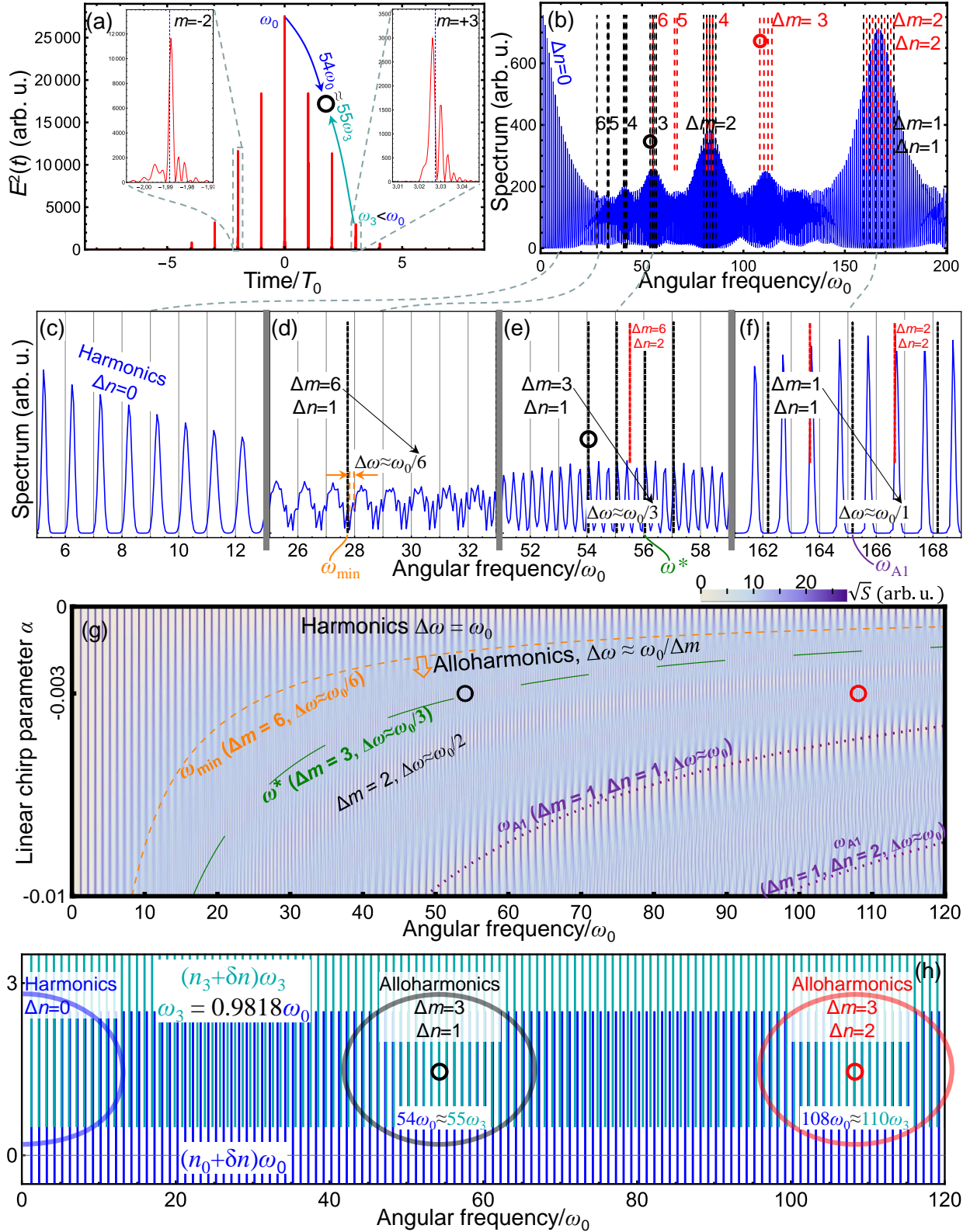

FIG. 2. Alloharmonic generation by a driver with linear chirp, model described in Appendix B [Eqs.(10),(15) with α=−0.003, δn=0.25, $n_{max}$=200, $a_n$=1, $\phi_n$=−[(n−100)/75]$^2$, and Gaussian envelope

$A_\tau(t)=\exp(-t^2/\tau^2)$, $\tau=3T_0$]. (a) Time domain; arrows indicate an example of ($\Delta n=1,\Delta m=3$) interharmonic interference of the $n=54^{th}$ harmonic of the central pulse ($m=0,\omega=\omega_0$) with the $n+1=55^{th}$ harmonic of pulse number $m=3$ with lower frequency $\omega_3<\omega_0$, Eq.(1). This interference, which is a part of the alloharmonic series [41] discussed in Appendix B [FIG. 7(a,b)], is indicated by black circles in (a,b,e,g,h), FIG. 7(c), and the Supplemental Movie [42]. Another interference example, ($n=108,\Delta n=2,m=0,\Delta m=3$), is indicated by red circles. Insets, $m=-2$ and $m=+3$ pulses in the time domain; vertical dashed lines show calculated pulse positions [solutions of Eq.(11)]; in the periodic case, these pulses would occur at $t=-2T_0$ and $t=3T_0$. (b) Overall spectrum. Long black and short red vertical dashed lines in (b) and (c-f) correspond to $\Delta n=1$ and $\Delta n=2$ alloharmonics [solutions of Eq.(1)], with the $\Delta m$ shown by the black ($\Delta n=1$) and red ($\Delta n=2$) numbers. (c-f) Spectra in limited bandwidth ranges: (c) Approximate harmonics, $\Delta n=0$, with fringe separations $\Delta\omega=\omega_0$, (d-f) alloharmonics with the main fringe separations $\Delta\omega\approx\omega_0/6$ ($\Delta m=6,\Delta n=1$), $\Delta\omega\approx\omega_0/3$ (mainly $\Delta m=3,\Delta n=1$), and $\Delta\omega\approx\omega_0/1$ (mainly $\Delta m=1,\Delta n=1$), respectively; vertical dashed lines show calculated interharmonic interferences, Eq.(1). (g) Variation in the spectral amplitude with the linear chirp parameter $\alpha$ (also [42]). The dashed orange and long-dashed green lines indicate $\omega_{min}$ (alloharmonic appearance) and $\omega^*$ (alloharmonic dominance), Eq.(4). Dotted lines, frequencies $\omega_{A1}$ for ($\Delta m=1,\Delta n=1$) and ($\Delta m=1,\Delta n=2$), Eq.(6), where the alloharmonic fringe separation is approximately $\omega_0$. (h) Moiré pattern of the harmonic combs of the $m=0$ (blue) and $m=3$ (cyan) pulses [arrows in (a)]. Ellipses, interference regions corresponding to the same-order-interference (harmonics, $\Delta n=0$), and interharmonic interferences (alloharmonics, $\Delta n=1$ and $\Delta n=2$).

### C. Alloharmonics in the BISER experiment

Alloharmonics explain the entire complex experimental spectrum, FIG. 1(a), i.e. different fringe periods with the $\approx$1:2:3 ratios, FIG. 1(e-g), and their positions, FIG. 1(b-d). To gain insight into the driver properties, we fitted the experimental spectrum with the alloharmonics model [FIG. 3(a) red line; details are in Appendix C, Table I middle column, and FIG. 9(a)]. Five parameters of the alloharmonic model, $\lambda_0$, $\alpha$, $\beta$, $\gamma$, and $\delta$, precisely described the complex experimental fringes. In contrast, a harmonic model with a constant frequency, Table I right column, did not fit the data well, FIG. 9(b); the Bayesian information criterion (BIC) decisively favored the alloharmonic model, with $\Delta$BIC~600.

The alloharmonic fit gave $\hbar\omega_0=0.63123\pm0.00010$ eV ($\lambda_0=1964.2\pm0.3$ nm), which is ~2.4 times lower than the laser frequency. We checked that frequencies close to the laser frequency ($\hbar\omega_L\approx1.53$ eV) gave a poorer fit with $\Delta$BIC>1000 (Appendix C). This sensitivity led to precise parameter determination and small fitting errors. The fitted dimensionless chirp parameters were relatively large, with the leading linear term being $\alpha=-(1.684\pm0.007)\times10^{-2}$. Figure 3(b) shows amplitudes and time-shifts of the five strongest pulses in the time domain. The significant shifts from the periodic case, $\delta t_m=t_m-mT_0>0.1T_0$ for $m=\pm2$, were consistent with the notable Fourier component shifts from $\Delta t/T_X=2$ in FIG. 1(e,f). This experiment and fit demonstrate that the alloharmonics allow precisely determining nonperiodic driver parameters, such as its central frequency and time variation, even for high chirp orders, up to $\delta$ (4$^{th}$ order) in our case.

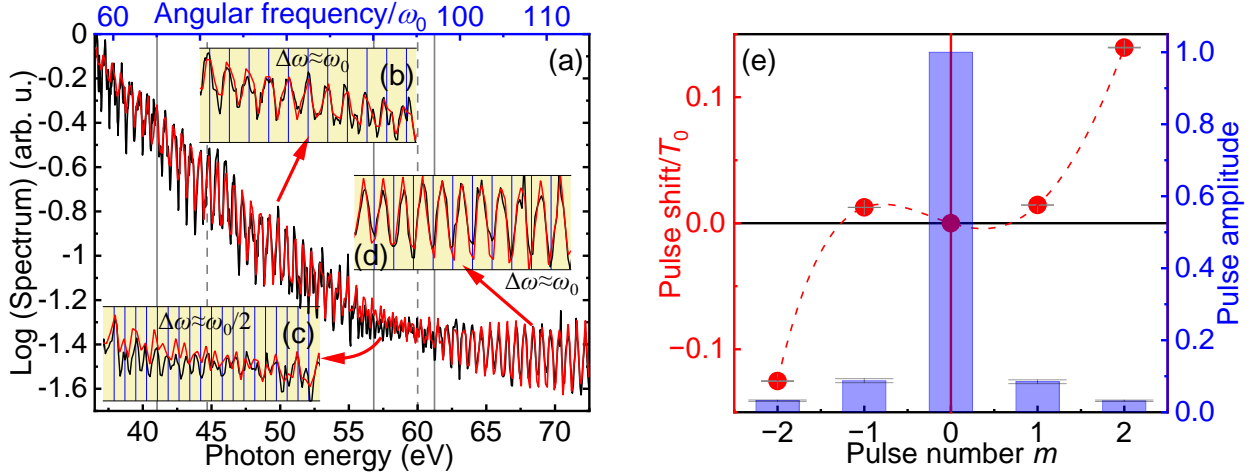

FIG. 3. Alloharmonic fit of the experimental BISER spectrum. (a) Black, the experimental BISER spectrum [same as FIG. 1(a) black]; red, the alloharmonic fit (Appendix C), with the fundamental photon energy $\hbar\omega_0$=0.63123 eV (wavelength $\lambda_0$=1964.2 nm) and other fit parameters shown in Table I, middle column. Grey vertical lines indicate the positions of alloharmonics with interharmonic separations $\Delta n$ from 1 to 4 and separations of pulses in the time domain with $\Delta m$=1 (solid) and $\Delta m$=2 (dashed), which explain the spectral fringes with spacings of $\Delta\hbar\omega \approx \hbar\omega_0/\Delta m = \hbar\omega_0$ in (b,d) and $\approx \hbar\omega_0/2$ in (c), respectively. (e) Five strongest time-domain pulses calculated from the alloharmonic model. Red circles (left axis) show the normalized shifts (residuals) from the periodic positions, $\delta t_m/T_0 = t_m/T_0 - m$, where the period $T_0 \approx 6.55$ fs corresponds to the fundamental photon energy $\hbar\omega_0$. The error bars (some are difficult to see) are calculated by the error propagation technique from the fit standard errors (Appendix C, Table I).

### D. Alloharmonics in gravitational waves emitted by binary black hole mergers

Binary black holes inspiral because they lose energy to gravitational waves. As the inspiral progresses, the orbital frequency gradually increases, thus providing nonperiodicity and therefore conditions for alloharmonics. Gravitational wave spectrometers are currently unavailable; thus, we used a gravitational waveform from published state-of-the-art numerical relativity calculations [43, 44] [red line in FIG. 4(a)] describing a sufficiently long (>50 cycles) evolution. The orbital frequency $\omega_{Orb}$ increases slowly at first but quite rapidly near the merger at $t \sim 24\,000M$ (blue line in FIG. 4(a); we use geometric units; here, $M$ is the sum of gravitating masses, Appendix D). We assume that a gravitational spectrometer would have a millisecond-range Gaussian time window, such as the two examples in FIG. 4(a): for a combined 20-solar-mass binary system, these window durations are ~100 ms. Results for other window durations are in the Supplemental Material [45].

The gravitational harmonics are well phased, as evidenced by the short gravitational wave bursts [FIG. 4(b)]. These bursts are analogous to the attosecond pulses generated by laser harmonics [12, 13]. The positions of these bursts are well described by Eq.(11), with the phase $\Phi(t)$ determined from integrating the orbital frequency dependence $\omega_{Orb}(t)$ [45].

The spectra in the two windows are shown in FIG. 4(c,d), and the spectrum evolution in the moving time window in the Supplemental Movie [46]. Initially, the windowed spectrum consists of normal harmonics because the orbital frequency is approximately constant, FIG. 4(c). However, as the inspiral progresses, the orbital frequency change accelerates, and within the same window duration, the frequency increases by 20% [45], thus alloharmonics emerge starting at $4\omega_{Orb}$, corresponding to interharmonic separations $\Delta n = -1$, i.e., interferences with order $n$ at time $t_m$ and order $n-1$ at time $t_{m+\Delta m} > t_m$, FIG. 4(d). These interferences [calculated with the alloharmonic equation, Eq.(1)] are shown by the dashed vertical lines in FIG. 4(d). The gravitational wave spectrum intensity decreases rapidly with frequency; hence, larger interharmonic separations ($\Delta n=-2$, etc.) are less pronounced. A moderate resolving power of $R_{min}\approx 29$ would be required to observe these gravitational wave alloharmonics.

The chirp rate in this example is orders of magnitude larger than that induced by the Earth's rotation and motion (Section II.F), and therefore, these alloharmonics can be observed without correction to the solar system's barycenter.

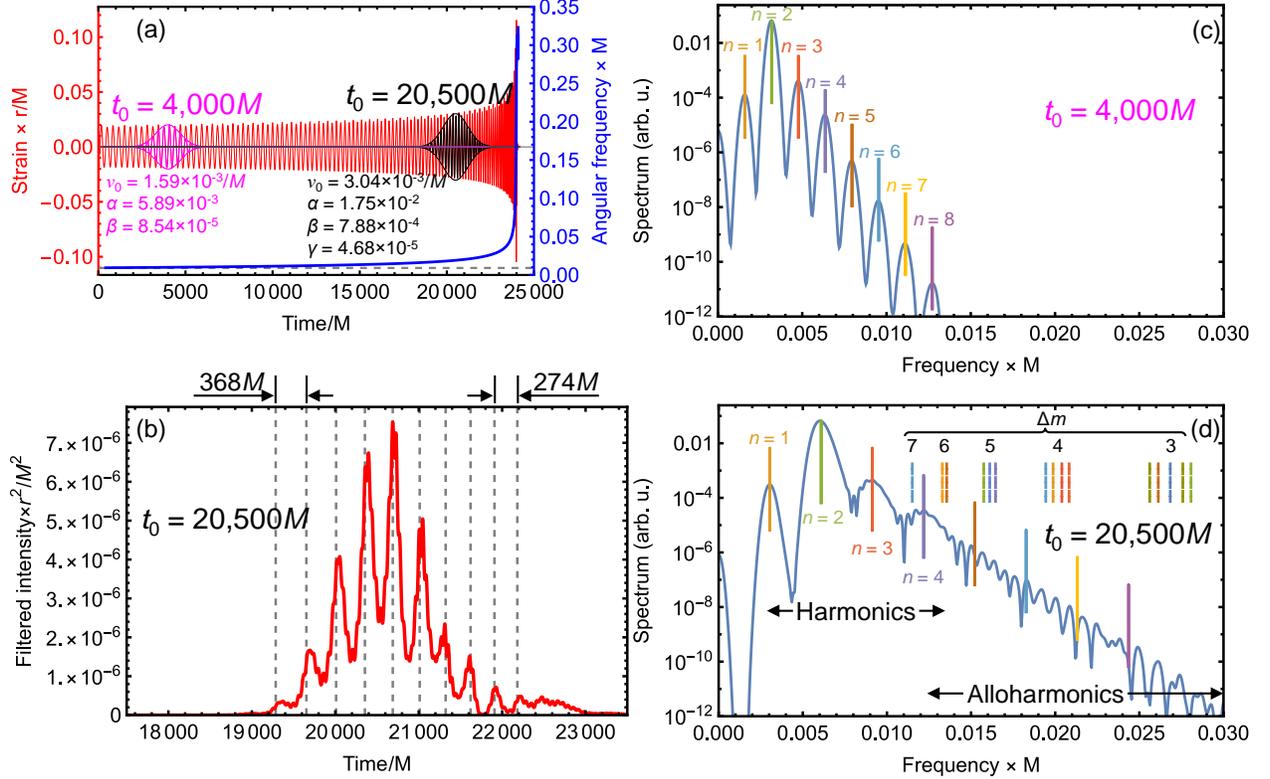

FIG. 4. Alloharmonics in a gravitational wave emitted by an inspiralling binary black hole. (a) Red (left axis), normalized strain vs time, numerical relativity calculation from the SXS Gravitational Waveform Database [44] SXS:BBH:1148, with a mass ratio of 2.038, dimensionless spins of 0.43 and 0.51, and initial orbital angular frequency of $\omega_{Orb,ini}=9.165\times 10^{-3}/M$. Magenta and black, windowed strains used to calculate spectra in (c,d) (Gaussian windows, durations $\tau=1000M$) and corresponding fitted frequency and chirp parameters. Blue (right axis), calculated [45] instant orbital frequency; horizontal dashed line is the initial orbital frequency $\omega_{Orb,ini}$. (b) Gravitational bursts produced by a summation of $n\geq 3$ harmonics in the black window in (a); vertical dashed lines, burst positions calculated by solving Eq.(11). (c,d) Spectra of strains in magenta and black windows in (a), respectively; solid vertical lines, harmonics of the instantaneous orbital frequency shown in (a). Dashed vertical lines, alloharmonics, i.e. solutions of Eq.(1) with $\Delta n=-1$ and $\Delta m$ shown above. The number of alloharmonic fringes between harmonics is approximately equal to the corresponding $\Delta m$. In all frames, $M$ is the sum of the two Christodoulou masses, i.e., gravitating masses including the rotational energies; the time unit is $GM/c^3$, the unit of distance $r$ is $GM/c^2$ (geometric units), where $G$ is the gravitational constant. For $M=M_\odot$ (one solar mass), these are 4.93 μs and 1.48 km, respectively.

### E. Alloharmonics in frequency combs

Optical frequency combs are spectra of highly-periodic pulse trains generated by a laser pulse bouncing in a stabilized cavity with a partially-transmissive output coupler. Frequency combs, which are composed of equally-spaced spectral lines separated by the train repetition rate, revolutionized precision spectroscopy by enabling fast self-calibrated measurements with atomic clock accuracy [5, 6, 47, 48] and temporal coherence up to 2000 s, Ref.[49].

Here we consider an optical comb from a slightly nonperiodic pulse train, generated by an in-vacuum cavity with its end mirror moving with velocity $V \ll c$; here $c$ is the speed of light in vacuum. The optical path between bounces is $2(L_0+Vt)$, where $L_0$ is the initial cavity length with $T_0=2L_0/c$ the initial round-trip period. The $m^{\text{th}}$ pulse in the train occurs at $t_m = \frac{L_0}{V}\left[\left(\frac{c+V}{c-V}\right)^m - 1\right] \approx T_0\left(m + \frac{V}{c}m^2 + \cdots\right)$, yielding the chirp parameter $\alpha = -V/c$, Supplemental Material [50]. Due to the Doppler shift, the alloharmonic equation (1) reads $(n+\delta n)\left(\frac{c-V}{c+V}\right)^m \omega_0 = (n+\delta n+\Delta n)\left(\frac{c-V}{c+V}\right)^{m+\Delta m}\omega_0$, yielding $n = \frac{\Delta n}{\left(\frac{c+V}{c-V}\right)^{\Delta m}-1} - \delta n$. For the observation time $\tau \approx \Delta m T_0$, the alloharmonics start from frequency $\nu_{\min} \approx \frac{c}{2\Delta m|V|T_0} = \frac{1}{2\Delta m|\alpha|T_0}$, Eq.(4) and [50], and the minimum required resolving power (for $|\Delta n|=1$) is $R_{\min} \approx \frac{c}{2|V|} = \frac{1}{2|\alpha|}$, Eq.(7). For a typical 150 cm cavity length (round-trip period $T_0 \approx 10$ ns, repetition rate $\nu_0 \approx 100$ MHz), velocity $V=1$ mm/s, and observation time $\tau=1$ ms ($\Delta m=\tau/T_0 \approx 10^5$ pulses, 1 μm mirror shift), these yield $\alpha \approx -3.34 \times 10^{-12}$, $R_{\min} \approx 1.5 \times 10^{11}$, and alloharmonics replace the normal comb in the entire near-infrared, visible, and ultraviolet regions ($\nu > 150$ THz, $\lambda < 2$ μm).

Alloharmonics in frequency combs allow, on the one hand, a precise velocity measurement and stabilization, and, on the other hand, a comb density control without replacing the cavity. For the precision frequency metrology [5, 6], denser combs provide a larger number of reference frequency points at more localized positions in the spectrum, facilitating higher accuracy; on the other hand, this requires higher resolving power and thus often needs to be avoided; for the latter cases, alloharmonics determine the maximum cavity stabilization velocity.

Alloharmonics in optical frequency combs allow precise measurements of extremely small accelerations. This provides a new method for testing gravity theories such as modified Newtonian dynamics [51], which requires sensitivity to accelerations of $a \sim 10^{-10}$ m/s². An accelerating cavity mirror induces a quadratic chirp (Appendix B Section E, FIG. 8), with $\beta=-(aT_0)/3c$, and thus, alloharmonics at frequencies $\nu > \frac{\nu_0}{3|\beta|\Delta m^2} = \frac{c}{a\tau^2}$ reveal acceleration, with the required resolving power $R > \frac{1}{3|\beta|\Delta m} = \frac{c}{a\tau}$. For a $\tau=200$ s observation time, these give $\nu>75$ THz ($\lambda<4$ μm) and $R \sim 10^{16}$; in contrast to linear chirp cases, here longer observation time reduces the required resolving power.

### F. Alloharmonics in radio emission of pulsars

Pulsars are spinning neutron stars emitting periodic electromagnetic pulses [52], similar to a lighthouse. Pulsar radio spectra are coherent, consisting of harmonics of the rotation frequency. Fourier transforms of the pulsar radiation intensity, i.e., their power spectra, contain interesting features [53]. The pulsar rotation decelerates gradually, providing conditions for alloharmonics. Some pulsars exhibit glitches [25], sudden rotation accelerations, which can provide information about pulsar structure and physics.

Pulsars have nearly-constant rotation frequencies $\nu_0$ with small time derivatives $\dot{\nu}$, producing the chirp parameter $\alpha \approx \dot{\nu}/2\nu_0^2$, Supplemental Material [54]. For an observation time window consisting of $N_p$ periods, the alloharmonics can be observed from frequencies $\nu_{\min} = -\nu_0^3/N_p\dot{\nu}$ with resolving power $R_{\min} = -\nu_0^2/\dot{\nu}$, independently on the observation time. In our examples, Appendix E (Table II) and Supplementary Materials [54], we assume a ~200 s observation time, which is typical of meridian instruments [53] rotating with the Earth; however, other instruments can have longer time windows.

The Vela pulsar (J0835-4510, age ~$1.1 \times 10^4$ years) has $\nu_0=11.1946499395$ Hz and $\dot{\nu}=-1.5666 \times 10^{-11}$ Hz² (Ref.[55]). These parameters yield $\alpha \approx -6.25 \times 10^{-14}$ and $N_p=2300$ periods in a 205 s time window. The minimum frequency above which alloharmonics appear is $\nu_{min}\approx38.9$ GHz, and the required resolving power is $R_{min}\approx8.0\times10^{12}$. For the older Geminga pulsar [56], $|\alpha|$ is smaller, and the alloharmonic frequency is higher, $\nu_{min}\approx445$ GHz. On the contrary, for young pulsars, the deceleration is much faster; for example, for SGR 1806-20 (J1808-2024, age ~220 years), $\nu_0=0.1323466$ Hz and $\dot{\nu}=-9.62\times10^{-12}$ Hz$^2$ (Ref.[55]), yielding $\alpha\approx-2.75\times10^{-10}$ and $N_p=27$ periods in a 204 s time window, with $\nu_{min}\approx8.92$ MHz and $R_{min}\approx1.8\times10^9$. Alloharmonics from the same pulsar can be observed in a considerably shorter window, with $N_p=2$ periods (~15 seconds, 3 pulses), $\nu_{min}\approx120$ MHz, and the same resolving power. Thus, generally, the spectra of young pulsars contain only alloharmonics within typical radio telescope bands, i.e., the atmosphere's radio window, ~20 MHz to 1 THz, while older pulsars have normal harmonics up to a few tens of GHz and alloharmonics at higher frequencies.

The chirp parameter for a glitch is $\alpha=\Delta\nu_G/4\pi N_G\nu_0$ [54], where $\Delta\nu_G$ and $N_G$ are the frequency change and number of periods during the glitch. The alloharmonics start from $\nu_{min}=2\pi\nu_0^2/\Delta\nu_G$ and their observation requires the resolving power $R_{min}=2\pi N_G\nu_0/\Delta\nu_G$ [54]. For example, the 2016 Vela glitch had $\Delta\nu_G/\nu_0=1.431\times10^{-6}$ and a duration of $N_G=49$ periods (4.4 seconds) [25]. For these, $\alpha=2.32\times10^{-9}$ and the alloharmonics start from $\nu_{min}\approx49.2$ MHz ($R_{min}\approx2.2\times10^8$).

Pulsars apparent nonperiodicity is also caused by the Earth rotation and orbit: detector motion along a circle with radius $r$ and angular frequency $\Omega$ induces the apparent chirp coefficient $\alpha\approx-r\Omega^2/2c\nu_0$ for an in-plane propagating plane wave (Table II, [54]). Despite this, glitch alloharmonics are observable, as glitches induce much larger chirp coefficients. Spin-down alloharmonics of the youngest pulsars are observable by spaceborne instruments orbiting the Sun. For older pulsars, a correction to the solar system's barycenter, or a special case of wave propagation perpendicular to the detector rotation plane, would be required.

## III. DISCUSSION

Reliable alloharmonic observations must satisfy two conditions: Sufficient resolving power to observe fine fringes, and observation of at least two different fringe periods. Counterintuitively, the same apply for harmonics, to distinguish them from possible alloharmonics. These conditions are challenging, especially at high harmonic orders [57, 58]. Whereas several high-quality experiments at lower orders, such as [17-19] and our data presented in Section II.A, satisfy these conditions, in other cases alloharmonics can easily escape detection and cause significant errors. First, insufficient resolving power leads to absence of expected harmonic fringes, which can be misinterpreted as absence of temporal coherence or even as a single peak ("isolated attosecond pulse") in the time domain, whereas a nonperiodic coherent pulse train produces much finer alloharmonic fringes that remain unresolved. Second, alloharmonics observed in a narrow bandwidth can be misinterpreted as harmonics with a wrong "driver frequency" and "carrier-envelope offset;" this is especially probable for alloharmonics with pulse separation $\Delta m=1$, or for the quadratic chirp case (Appendix B Section E), for both of which the spectral fringe period, $\Delta\omega\approx\omega_0/\Delta m$, is similar (but not exactly equal) to the expected driver frequency $\omega_0$.

In sufficiently nonperiodic cases, alloharmonics may appear even at very low frequencies, as in the gravitational wave emitted by an inspiralling binary black hole near merger, FIG. 4(d) (around harmonic order $n\sim4$), or in the Sliding Mirror (allo)harmonics, Fig. 1 of Ref. [59] ($n\sim5$).

Observation of alloharmonics in optical frequency combs and pulsar spectra will require high resolving powers which are not practical with standard spectrometers (as they would have light-minute sizes), but can be obtained with dual-comb spectroscopy [6, 60].

Analytical formulas, Eq.(19), Ref.[41], used to calculate interharmonic interference in FIG. 7(b), can be used to calculate very-high-frequency spectra, for which the numerical Fourier transform is time consuming [54].

Alloharmonics from the BISER experiment, Table I middle column, and black hole merger, FIG. 4(a), have similar absolute values of the dimensionless chirp |α|, with opposite signs. However, the spectrum is insensitive to the time inversion. Thus, if only spectral shapes and fringes are important, this provides a basis for scaled, i.e., with the same dimensionless parameters, laboratory astrophysics experiments, where gradually decreasing experimental frequencies can be used to model both processes with decreasing and increasing frequencies.

## IV. CONCLUSION

We developed a theory of *harmonic* and *alloharmonic* spectra formation in nonperiodic cases, where different harmonic orders start to interfere at higher frequencies. This interference produces additional, much finer, spectral fringes (*alloharmonics*). Our basic equations are valid for an arbitrary degree of nonperiodicity. We also examined in detail the important case of small nonperiodicity and showed that in this case the normal harmonic fringes with gradually increasing bandwidth are formed at low harmonic orders, while, at higher frequencies, the alloharmonics have spectral fringe separations approximately equal to various integer *fractions* of the base (driver) frequency. With this theory, we predicted the frequencies at which alloharmonics appear in the spectra, their fringe spacing, and required resolving power and bandwidth to correctly measure these fringes. The theory explained our experimental data and explained why they were inexplainable without it. The theory also allows reconstruction of the parameters of driver nonperiodicity from measured spectra, which we did for our experimental BISER data. The theory is widely applicable to all nonperiodic harmonic-generating drivers. We analyzed gravitational wave spectra from previously published state-of-the-art numerical relativity simulations[43, 44] of inspiralling binary black holes and obtained results consistent with our theory. We also predicted important implications for frequency comb spectroscopy, where the alloharmonic concept allows extension into regimes with moving and accelerating laser cavity mirrors, and pulsar studies, in both of which previously unavailable nonperiodicity properties can be derived from alloharmonic observations. We argued that unaccounted-for alloharmonics can cause a lot of confusion and misinterpretation of observations in laser labs and astrophysics.

Alloharmonic theory fills in a gap of knowledge related to present day laser-driven harmonic experiments. Furthermore, it predicts yet-to-be-observed alloharmonic spectra in optical frequency combs and astrophysics. Like our rich experimental spectra, these observations will allow inferring the driver periodicity and nonperiodicity parameters. However, we expect that, somewhat counterintuitively, these spectra will be most interesting and informative by their dissimilarities with the "simple" alloharmonics due to such effects as higher chirp orders, possible reduction in the coherence degree and pulsar nullings [25], binary black hole inspiral in the presence of other objects, and other yet-to-be-discovered factors which will require further research and expanded models.

The experimental data are available from the corresponding author on reasonable request. The code written in the Wolfram Language is published in the Supplemental Materials [41, 45, 50, 54]. The gravitational waveforms used in this study are available from the Simulating eXtreme Spacetimes (SXS) Collaboration catalogue [43, 44].

## ACKNOWLEDGEMENTS

We are grateful to the J-KAREN-P laser team and acknowledge discussions with Dr. V. M. Malofeev and Dr. S. B. Echmaev. This research was supported by JSPS Kakenhi JP26707031, JP19H00669,

JP19KK0355, and JP23H01151, and Strategic Grants by the QST President: Creative Research #20 and IRI.

A.S.P. led the research, K.O., A.S., A.Ya.F., T.A.P., H.Ko., Y.H., Y.F., H.D., N.H., M.I., M.N., M.K., T.K., H.Ki., M.K., D.N., and A.S.P. prepared and performed the experiment, M.S.P. proposed the alloharmonic explanation of the data, M.S.P., T.Zh.E., J.K.K., S.V.B., and A.S.P. developed the theory and applied it to the BISER spectra, gravitational waves, frequency combs, and pulsar radio emission. M.S.P., T.Zh.E., J.K.K., and A.S.P. prepared the manuscript with contribution from other authors. All authors contributed ideas.

The authors declare no competing interest.

# APPENDIX A: EXPERIMENT ON COHERENT EXTREME ULTRAVIOLET GENERATION VIA THE BISER MECHANISM

### A. Laser

We used the multi-terawatt femtosecond linearly polarized J-KAREN-P laser [61] based on chirped pulse amplification technology [20]. The laser pulses had a central wavelength of $\lambda_L=0.8115\pm0.0007$ μm, corresponding to the photon energy of $\hbar\omega_L=1.5278\pm0.0013$ eV; here and throughout this paragraph, the errors listed are the standard deviation of the shot-to-shot parameter variations. The pulse energy was $\mathcal{E}_L=0.648\pm0.011$ J. The pulse duration and spot shape were measured in vacuum at full laser power attenuated with two wedges and additional high-quality neutral-density filters in the case of the spot measurement. The pulse measured with self-referenced spectral interferometry [62] had a full width at half maximum (FWHM) duration of 33±5 fs and an effective duration of $\tau_{Eff}=38\pm6$ fs. Here, $\tau_{Eff}=\int P_n(t)dt$ is the area under the normalized power curve $P_n(t)$, Ref.[63]. The peak power $P_L=\mathcal{E}_L/\tau_{Eff}$ was 17±3 TW. The pulses were focused with an f/9 off-axis parabolic mirror. The focal spot, measured with an achromatic lens and high dynamic range charge-coupled device (CCD), had an FWHM of (10.1±0.2) μm × (12.1±0.5) μm, horizontal (along the polarization direction) × vertical, and an effective spot radius [63] of $r_{Eff}=8.3\pm0.3$ μm. The estimated irradiance in the absence of plasma was $I_{L,Vac}=P_L/\pi r_{Eff}^2=(7.9\pm1.4)\times10^{18}$ W/cm$^2$, and the corresponding dimensionless amplitude was $a_{0,Vac}=1.9\pm0.2$. In plasma, these values increased significantly due to relativistic self-focusing [64-67] (see below the 'Laser propagation in relativistic plasma' section).

### B. Gas jet

We used a 1-mm diameter supersonic helium gas jet with a Mach number of 3.3, as described in [68]. The peak electron density in the helium jet, which is controlled by the backing pressure, was set to $n_e\approx(2.6\pm0.8)\times10^{19}$ cm$^{-3}$. Here, the 30% error is due to interferometry reconstruction errors.

### C. Relativistic plasma

In relativistic plasma, the electron velocity is close to the speed of light in vacuum, $c$. The electron dynamics in the laser field $E_L$ is relativistic when the laser dimensionless amplitude, $a_0=eE_L/m_e c\omega_L=(I_L/I_R)^{1/2}$, is comparable to or greater than unity [69]. A laser is relativistically strong when $a_0\geq1$, as in our experiment. Here, $e$ is the elementary charge, $m_e$ is the electron mass, $I_L$ is the peak irradiance, and $I_R=\pi m_e^2 c^5/2e^2\lambda_L^2\approx1.37\times10^{18}$ W/cm$^2\times(\lambda_L[\mu m])^{-2}$ for a linearly polarized laser with angular frequency $\omega_L$ and wavelength $\lambda_L$. $\lambda_L[\mu m]$ denotes the laser wavelength in micrometers. For $\lambda_L=0.81$ μm, we have $I_R\approx2\times10^{18}$ W/cm$^2$.

### D. Laser propagation in relativistic plasma

In our case, the plasma was underdense (in the small amplitude limit [70]); thus, its electron density $n_e \approx 2.6 \times 10^{19}$ cm$^{-3}$ was much smaller than the critical density, $n_{cr} = m_e\omega_L^2/4\pi e^2 \approx 1.1 \times 10^{21}$ cm$^{-3}/\lambda_L[\mu m]^2 \approx 1.7 \times 10^{21}$ cm$^{-3}$. Thus, the laser pulse propagated through the plasma. Furthermore, the peak power $P_L = 17 \pm 3$ TW significantly exceeded the threshold $P_{SF} = P_c(n_{cr}/n_e) \approx 0.6$ TW; thus, the laser underwent relativistic self-focusing [64-66]. Here, $P_c = (2m_e^2 c^5)/e^2 \approx 0.017$ TW, Ref. [66]. In the stationary self-focusing regime [71], the self-focused spot diameter and dimensionless amplitude are $d_{SF} = \lambda_L(a_0 n_{cr}/n_e)^{1/2}/\pi$ and $a_{SF} = (8\pi P_L n_e/P_c n_{cr})^{1/3}$, respectively. In our case, the estimated dimensionless amplitude after self-focusing was $a_{SF} \approx 7$.

A relativistically strong laser pulse propagating in underdense plasma excites wake waves such as longitudinal Langmuir waves [72]. Moreover, when the focal spot is narrower than the threshold $d_{BW} = 2\lambda_L(a_0 n_{cr}/n_e)^{1/2}/\pi$, bow waves are generated [73-75]. After relativistic self-focusing, this condition is always satisfied, as $d_{SF} < d_{BW}$, thus enabling excitation of prominent bow waves.

### E. BISER mechanism

A relativistic irradiance laser pulse (peak irradiance $I_0 > 10^{18}$ W/cm$^2$, dimensionless amplitude $a_0 \geq 1$) propagating in underdense plasma (typical electron density $n_e \sim 10^{19}$ cm$^{-3}$) generates plasma waves, significantly distorting the electron phase space and producing relativistic multistream plasma flow. In particular, phase space distortions corresponding to plasma waves such as wake waves [72, 76] and bow waves [73-75] can be observed (see the previous Section). In terms of catastrophe theory [77, 78], these distortions correspond to *fold singularities* and thus form sharp, high-density surfaces at the plasma cavity wall and bow wave front. At the joining of these two surfaces, there forms a circular line of higher-order *cusp singularity* characterized by even higher density and sharpness. For a linearly polarized laser, this circular line splits into two sharp high-density crescents. The electrons driven by the strong laser and plasma fields emit high-frequency radiation, which adds coherently within these singularity regions due to their high density and sharpness. For high frequencies, the coherence regions are small; thus, the BISER radiation is emitted from point-like (nanoscale) sources [4]. Catastrophe theory establishes that (i) the presence of multistream flows guarantees the formation of singularities, and (ii) the folds and cusps are structurally stable singularities, i.e., they do not disappear even under strong perturbations. Higher-order singularities may also form; however, they are not structurally stable. The BISER source generates extremely bright spatially and temporally coherent attosecond pulses that scale with the laser power; thus, the BISER source is considered a promising temporally and spatially coherent attosecond X-ray source [3, 4, 68, 79].

### F. BISER spectrum measurement

We used a grazing-incidence flat-field spectrograph [68] placed behind the gas jet (in the laser propagation direction). The spectrograph was composed of a gold-coated toroidal collecting mirror, a 200 µm slit, two 0.2 µm Al optical blocking filters, a spherical gold-coated varied line space (VLS) grating [80] with a central line density of 1200 lines/mm, and a back-illuminated CCD. A 0.4 T, 5-cm-long magnet deflected charged particles away from the spectrograph. In contrast to the previous experiments [3, 68], the spectrograph was operated in a longer wavelength region of 17-34 nm, which was determined by the L$_{2,3}$ absorption edge of the Al filters at 17 nm and its second diffraction order at 34 nm. In this spectral region, the achievable resolving power was much higher than that in the previous experiments, which allowed observation of alloharmonics with considerably finer fringes than normal harmonics. The operating bandwidth was wide enough to observe alloharmonics with several $\Delta m$ values, e.g., $\Delta m = 1$, 2 and 3 for the parameters in FIG. 1.

The spectrograph was carefully calibrated *in place* using the spectral lines of hydrogen-like helium and the position of the aluminum L$_{2,3}$ absorption edge, FIG. 5(a,b) and FIG. 6(a). The true wavelengths of the He lines, namely, the ordinates shown in FIG. 6(b), were taken from the NIST Atomic Spectra Database [81], and the Al absorption edge from the X-Ray Interactions With Matter database [82, 83]. Each spectral line in the recorded helium spectrum was fitted with a Gaussian profile, and its center was used as the corresponding abscissa. Thus, the line positions had subpixel accuracy. The dependence was fitted with a 3$^{rd}$-order polynomial using OriginLab's OriginPro software [84] with the default settings, FIG. 6(b), which provided an estimated accuracy of a few to several picometers, FIG. 6(c).

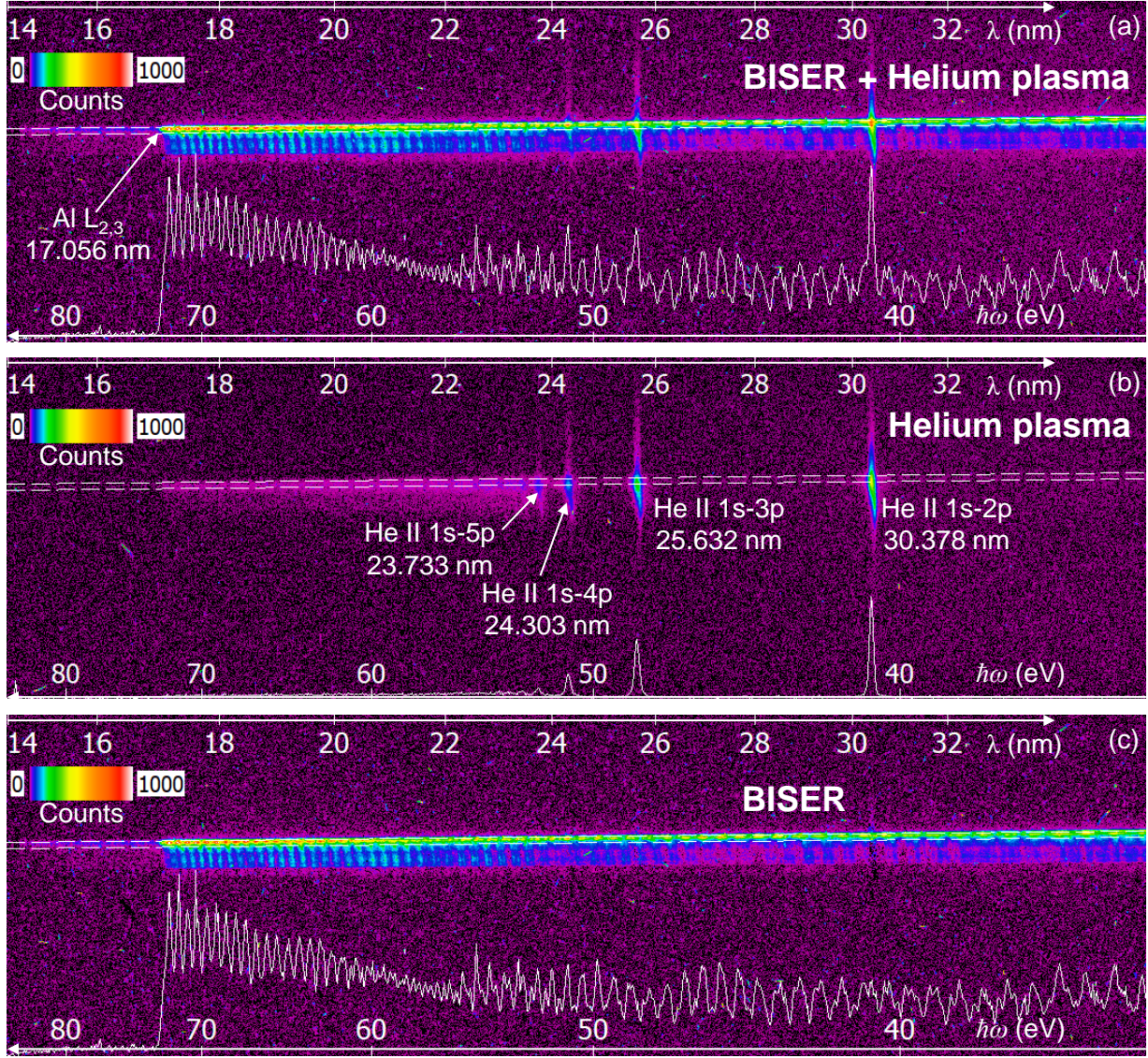

FIG. 5. Raw data recorded by the spectrograph. (a) Raw data with the combined BISER and helium plasma spectra. (b) Helium plasma spectrum. (c) BISER spectrum obtained by subtracting (b) from (a). Color scales show charge-coupled device (CCD) counts; lineouts correspond to the integrated counts between the dashed lines; the top and bottom axes correspond to the wavelength and photon energy, respectively.

The resolving power of the spectrograph was estimated from the BISER spectra using the finest observed fringes, red circles in FIG. 6(d). At wavelengths shorter than 21 nm, the resolving power was limited by the pixel size, while at longer wavelengths, the resolving power was limited either by geometric aberrations or by the finest fringe spacing in the experiment. The resolving power was much higher than that required to observe normal harmonics of the initial laser frequency [black dashed line in FIG. 6(d)], and two times higher than that required to observe alloharmonics in FIG. 1, i.e. alloharmonics with the redshifted frequency $\lambda_0$=1964 nm and $\Delta m$ up to 3, green dashed line in FIG. 6(d).

The raw spectra contained contributions from both the BISER source and the He plasma, FIG. 5(a), but the high reproducibility allowed the He contribution to be subtracted, FIG. 5(c). These BISER-only spectra were used in the analysis.

The raw spectra also contained a few pixels with high counts caused by hard x-rays. These pixels were identified and removed by an algorithm described in Appendix A of [85]. Due to high BISER brightness, this did not affect the experimental data quality.

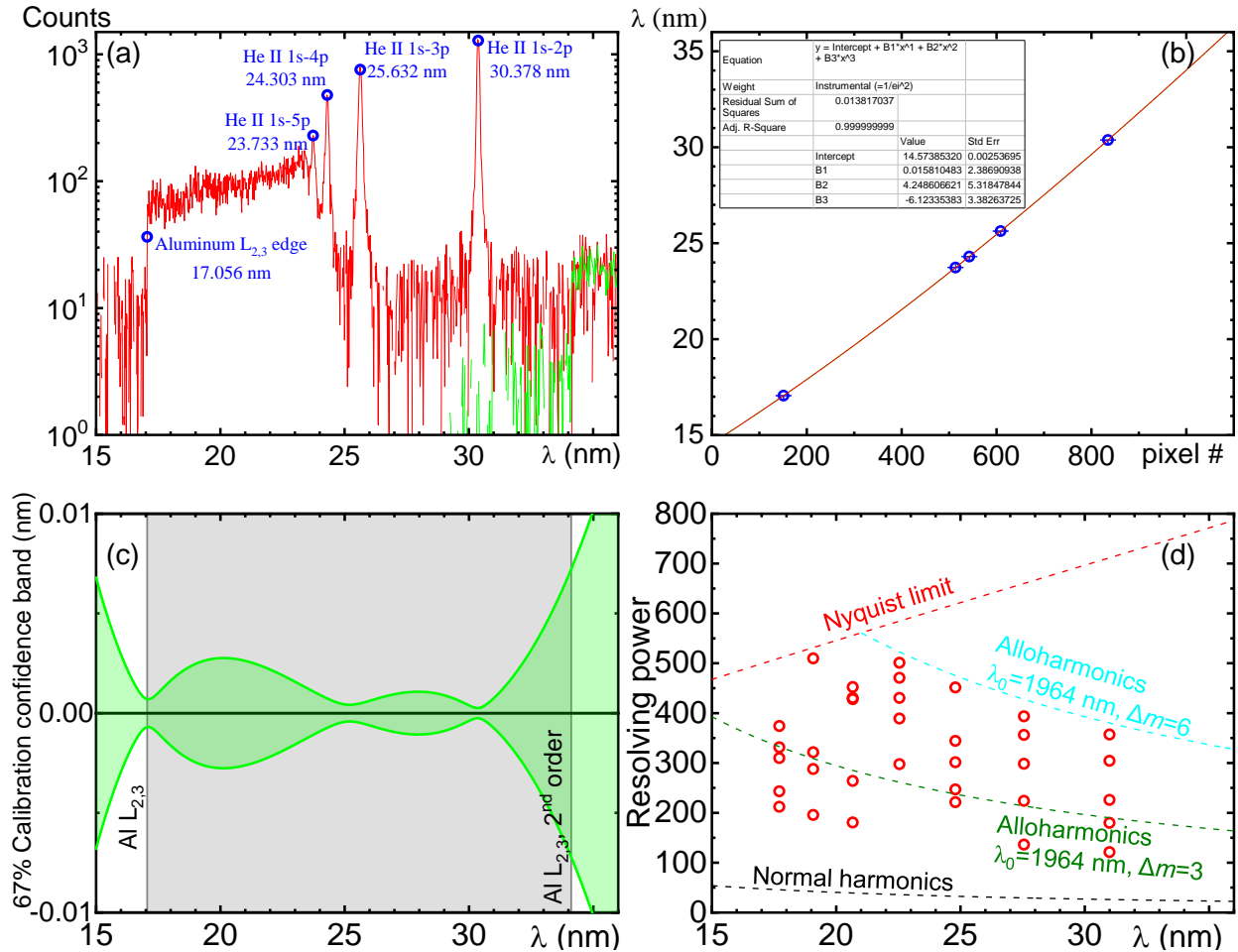

FIG. 6. Spectrograph calibration and resolving power. (a) Helium plasma spectrum excited in place by the same laser and recorded with the same spectrograph settings. Blue circles indicate spectral lines and the aluminum $L_{2,3}$ absorption edge used for wavelength calibration. (b) 3rd-order polynomial wavelength calibration (red line) through the He lines and Al $L_{2,3}$ absorption edge (blue circles; the hard-to-see error bars are the standard errors). (c) Estimated 67% confidence band of the wavelength calibration fit; the grey shaded area shows the spectrograph operating band between the Al $L_{2,3}$ absorption edge and its second diffraction order. (d) Resolving power of the spectrograph estimated from fine BISER alloharmonic fringes observed under various conditions (red circles). The red dashed line denotes the resolving power limit corresponding to two CCD pixels, the lower black dashed line represents the resolving power required to resolve normal harmonics with initial laser wavelength $\lambda_0$=0.81 μm, and the green and cyan dashed lines

represent the resolving powers required to resolve alloharmonics with red-shifted wavelength $\lambda_0$=1.964 µm and $\Delta m$=3 (as in FIG. 1) and $\Delta m$=6, respectively.

# APPENDIX B: DERIVATION OF EQUATIONS AND THEIR APPROXIMATE SOLUTIONS

## A. Alloharmonics and approximate harmonic solution

For non-constant frequency ($\omega_m \neq \omega_{m+\Delta m}$) and $\Delta n \neq 0$, the alloharmonic equation (1) has the following general solution for the order $n$ and alloharmonic frequency $\omega_A$:

$$(8)\quad n = \frac{\Delta n \omega_{m+\Delta m}}{\omega_m - \omega_{m+\Delta m}} - \delta n \approx \frac{\Delta n \omega_0}{\omega_m - \omega_{m+\Delta m}}$$

$$(9)\quad \omega_A = (n + \delta n)\omega_m = \frac{\Delta n \omega_m \omega_{m+\Delta m}}{\omega_m - \omega_{m+\Delta m}} \approx \frac{\Delta n \omega_0^2}{\omega_m - \omega_{m+\Delta m}}.$$

These equalities are valid for any, even significant, degree of nonperiodicity. In the approximate equalities on the right, we neglected $\delta n$ and used the small nonperiodicity approximation: $\omega_{m+\Delta m} \approx \omega_0$ and $\omega_m \omega_{m+\Delta m} \approx \omega_0^2$.

For slow frequency variation, $\omega_m \approx \omega_{m+\Delta m}$, and low orders $n$, the difference $(n+\delta n)(\omega_m - \omega_{m+\Delta m})$ is small. Thus, Eq.(1) is also *approximately* satisfied for $\Delta n$=0, representing the approximate harmonic solution. It is this approximate solution which appears as "harmonics" in all practical cases when the driver is almost periodic, i.e. its frequency is almost constant. At sufficiently high orders, Eq.(4), this frequency difference becomes large anyway and the harmonics vanish, while the alloharmonic solution always remains, with $\Delta n \neq 0$.

More than two equalities in the alloharmonic equation (1) can hold simultaneously at approximately the same frequency: the spectrogram in FIG. 7(a) shows how three harmonic orders, $n$=54, 55, and 56, from pulses with numbers $m$=−3, 0, and +3 (thus, two pairs with $\Delta m$=3 each), have nearly the same frequencies and therefore interfere producing fringes at $\omega \sim 55\omega_0$, with fringe spacing $\Delta\omega \approx \omega_0/3$. Indeed, the local portion of the spectrum around this frequency can be well represented by a sum of these three harmonics, FIG. 7(b).

## B. Alloharmonics model with nonlinear temporal phase

The Fourier transform of the spectrum, such as shown in FIG. 1(e-g), does not include phase and thus cannot reconstruct the true pulse train in the time domain – in particular, while the existing pulse separations can be obtained, their relative positions cannot. A model of high-order harmonics including temporal phase has been used, for example, in Ref.[47]. Its generalization to a nonconstant-driver-frequency case is:

$$(10)\quad E(t) = A_\tau(t) \sum_n a_n \cos[(n + \delta n)\Phi(t) + \phi_n].$$

Here, $E(t)$ is the amplitude (electromagnetic field, gravitational strain, etc.), $t$ is the time, $A_\tau(t)$ is the envelope with duration $\tau$ (assumed to be independent on the harmonic order, but it can be easily generalized to order-dependent durations), $a_n$ is the $n^{\text{th}}$ harmonic amplitude, $\phi_n$ is the spectral phase characterizing phase matching, $\Phi(t)$ is the driver temporal phase, $\omega(t) = \frac{d\Phi(t)}{dt}$ is the instantaneous driver frequency. The model and derivation of equations are in the Supplemental Material [41]. According to the known property of this model, for a reasonably small phase mismatch $\Delta\phi_n = \phi_{n+1} - \phi_n$, the amplitude $E(t)$ is close to zero everywhere except for several sharp pulses, FIG. 2(a), with the positions $t_m$ satisfying

$$(11)\quad \Phi(t_m) = 2\pi m - (\phi_{n+1} - \phi_n) \approx 2\pi m,$$

where $m$ is the pulse number: $m=0$ is the (main) pulse at $t=0$, while $m=\pm 1, \pm 2,...$ are (satellites) before ($m<0$) and after ($m>0$) this pulse. Periodic drivers, $\omega(t)=\omega_0$, produce equidistant pulses at $t_m=mT_0$ and equidistant harmonics with frequency separations of $\Delta\omega=2\pi/T_0=\omega_0$ due to the interference between the equal frequencies $(n+\delta n)\omega_0$ of all the pulses. For nonperiodic drivers, solution of Eq. (11) gives a nonperiodic pulse train, i.e., nonequidistant pulses in the time domain, with the frequency depending on the pulse number: $\omega_m=\omega(t_m)$. The small nonperiodicity approximation in the time domain, in the lowest order, is $t_m \approx mT_0$.

The alloharmonic concept allows exploring previously unknown properties of the model. The alloharmonic orders $n$ and frequencies $\omega_A=(n+\delta n)\omega_m$ are defined by the alloharmonic equation (1) for various integer values of $m$, $\Delta m>0$, and $\Delta n \neq 0$. The minimum orders $n_{min}$ and frequencies $\omega_{min}$ when the alloharmonics appear correspond to the minimum interharmonic separation, $|\Delta n|=1$ ($\Delta n=+1$ for decreasing and $\Delta n=-1$ for increasing frequency), and maximum interpulse separation $\Delta m_{max}$ producing noticeable fringes. At frequencies lower than $\omega_{min}$, the harmonics can be considered "normal", although the transition is gradual. For pulses providing noticeable fringes we use a condition that these pulses have a 1/e amplitude ($1/e^2$ intensity) with respect to the main ($t=0$, $m=0$) pulse. For a Gaussian amplitude envelope $A_\tau(t)=\exp(-(t/\tau)^2)$, this gives pulses at $t \approx \pm\tau$ with the separation $2\tau$ and $\Delta m_{max} \approx 2\tau/T_0$. Indeed, in our example with the duration $\tau=3T_0$ (three periods), fringes with the separation $\Delta\omega_{min} \approx \omega_0/6$ start to appear at $\omega_{min} \sim 28\omega_0$, corresponding to $\Delta m_{max}=6$, FIG. 2(d). For pulses producing dominant alloharmonics, we use a condition of $1/e^{1/2}$ intensity. For the Gaussian envelope this gives pulses at $t \approx \pm\tau/2$ with the separation $\tau$ and $\Delta m^* \approx \tau/T_0$. This is also visible in FIG. 2(e): alloharmonics start to dominate at $\omega^* \sim 56\omega_0$ with $\Delta\omega^* \approx \omega_0/3$. We note that these conditions are somewhat arbitrary and, in any case, approximate, as there may be no pulses exactly at the times $\pm\tau$ and $\pm\tau/2$. Further, these estimates are envelope-shape-dependent and detector (spectrometer) dependent: a higher signal-to-noise spectrometer can detect alloharmonics at lower frequencies.

The alloharmonic fringe spacing is

$$(12) \quad \Delta\omega = \frac{2\pi}{t_{m+\Delta m}-t_m} \approx \frac{2\pi}{(m+\Delta m)T_0-mT_0} = \frac{2\pi}{\Delta m T_0} = \frac{\omega_0}{\Delta m}.$$

Here the exact equality is valid for any degree of nonperiodicity, while in the approximate equality we used the small nonperiodicity approximation. We note that Eq.(12) is fully consistent with our experimental data, containing fringes with separations $\Delta\omega \approx \omega_x$, $\approx \omega_x/2$, and $\approx \omega_x/3$, FIG. 1(e-g), and with FIG. 2(d-f): fringes with $\Delta\omega \approx \omega_x$, $\approx \omega_x/3$, and $\approx \omega_x/6$. The fringe separation does not depend on the spectral phase $\phi_n$, although the time domain pulse shapes do.

The resolving power required to observe the alloharmonic fringes is defined as

$$(13) \quad R = \frac{\omega_A}{\Delta\omega} = \frac{(n+\delta n)\omega_m}{\Delta\omega} \approx \frac{n\omega_m}{\omega_0}\Delta m \approx n\Delta m,$$

i.e. $\sim\Delta m$ times larger than the resolving power required for harmonics at the same frequency, $R_H=n$.

### C. Alloharmonics model with polynomial phase

In many important cases the nonlinear driver phase can be written or approximated as a polynomial

$$(14) \quad \Phi(t) = \omega_0 t\left(1 + \alpha\frac{t}{T_0} + \beta\left(\frac{t}{T_0}\right)^2 + \gamma\left(\frac{t}{T_0}\right)^3 + \delta\left(\frac{t}{T_0}\right)^4 + \cdots\right)$$

with dimensionless nonperiodicity (chirp) parameters $\alpha$, $\beta$, $\gamma$, $\delta$, …. In the small nonperiodicity case these chirp parameters are much smaller than unity. The time derivative of the phase (14) gives the time-dependent frequency (2). Equation (11) with the phase (14) is a polynomial equation with solutions $t_m$ corresponding to the $m^{th}$ pulse positions. For the smallest polynomial orders, these solutions can be found analytically, although the explicit expressions may be cumbersome. Vice versa, given the pulse positions $t_m$, the chirp coefficients can be calculated from Eqs.(11),(14).

## D. Alloharmonics with a linear chirp (quadratic phase)

For the simplest linear chirp case

$$(15) \quad \Phi(t) = \omega_0 t \left(1 + \alpha \frac{t}{T_0}\right),$$
$$\omega(t) = \omega_0 \left(1 + 2\alpha \frac{t}{T_0}\right),$$

the nonperiodic pulse train has individual pulses at $t_m = \frac{\sqrt{1+4m\alpha}-1}{2\alpha} T_0 \approx (m - \alpha m^2) T_0$ with frequencies $\omega_m = \omega(t_m) = \omega_0 \sqrt{1 + 4m\alpha} \approx (1 + 2m\alpha)\omega_0$.

The alloharmonic orders are found from Eq.(1), which in this case is $(n + \delta n)\omega_0\sqrt{1 + 4m\alpha} = (n + \delta n + \Delta n)\omega_0\sqrt{1 + 4(m + \Delta m)\alpha}$. The alloharmonic frequencies are

$$(16) \quad \omega_A = (n + \delta n)\omega_m = -\frac{\Delta n \sqrt{1+4m\alpha}\left(1+4(m+\Delta m)\alpha+\sqrt{(1+4(m+\Delta m)\alpha)(1+4m\alpha)}\right)}{4\Delta m \alpha} \omega_0.$$

For $\alpha \ll 1$, this gives (5).

The alloharmonic fringe separations (12) are

$$(17) \quad \Delta\omega = \frac{2\pi}{t_{m+\Delta m} - t_m} = \frac{4\pi\alpha}{T_0\left(\sqrt{1+4(m+\Delta m)\alpha}-\sqrt{1+4m\alpha}\right)} \approx \frac{\omega_0}{\Delta m} + \left(1 + \frac{2m}{\Delta m}\right)\alpha\omega_0.$$

The required resolving power (13) is

$$(18) \quad R = \frac{\omega_A}{\Delta\omega} = -\frac{\Delta n \omega_m \omega_{m+\Delta m}}{2\omega_0^2 \alpha} \approx -\frac{\Delta n}{2\alpha}.$$

For very high frequencies and harmonic orders, such as those in radio spectra of pulsars and in optical frequency combs, numerical Fourier transforms can be very time-consuming. In the case of a Gaussian envelope and quadratic phase, the spectrum $S(\nu)$ can be calculated analytically [54]:

$$(19) \quad E(\nu) = \sum_n \frac{a_n}{2\sqrt{2\left(\frac{T_0}{\tau}\right)^2 + i4\pi n\alpha}} \mathrm{Exp}\left(-\frac{\left(\frac{\nu}{\nu_0}-n\right)^2 \pi^2}{\left(\frac{T_0}{\tau}\right)^2 + i2\pi n\alpha}\right),$$

$$S(\nu) = |E(\nu)|^2.$$

This formula was used to calculate interference of three and two harmonics, red and green lines in FIG. 7(b), and model-based pulsar spectra in the Supplemental Material [54].

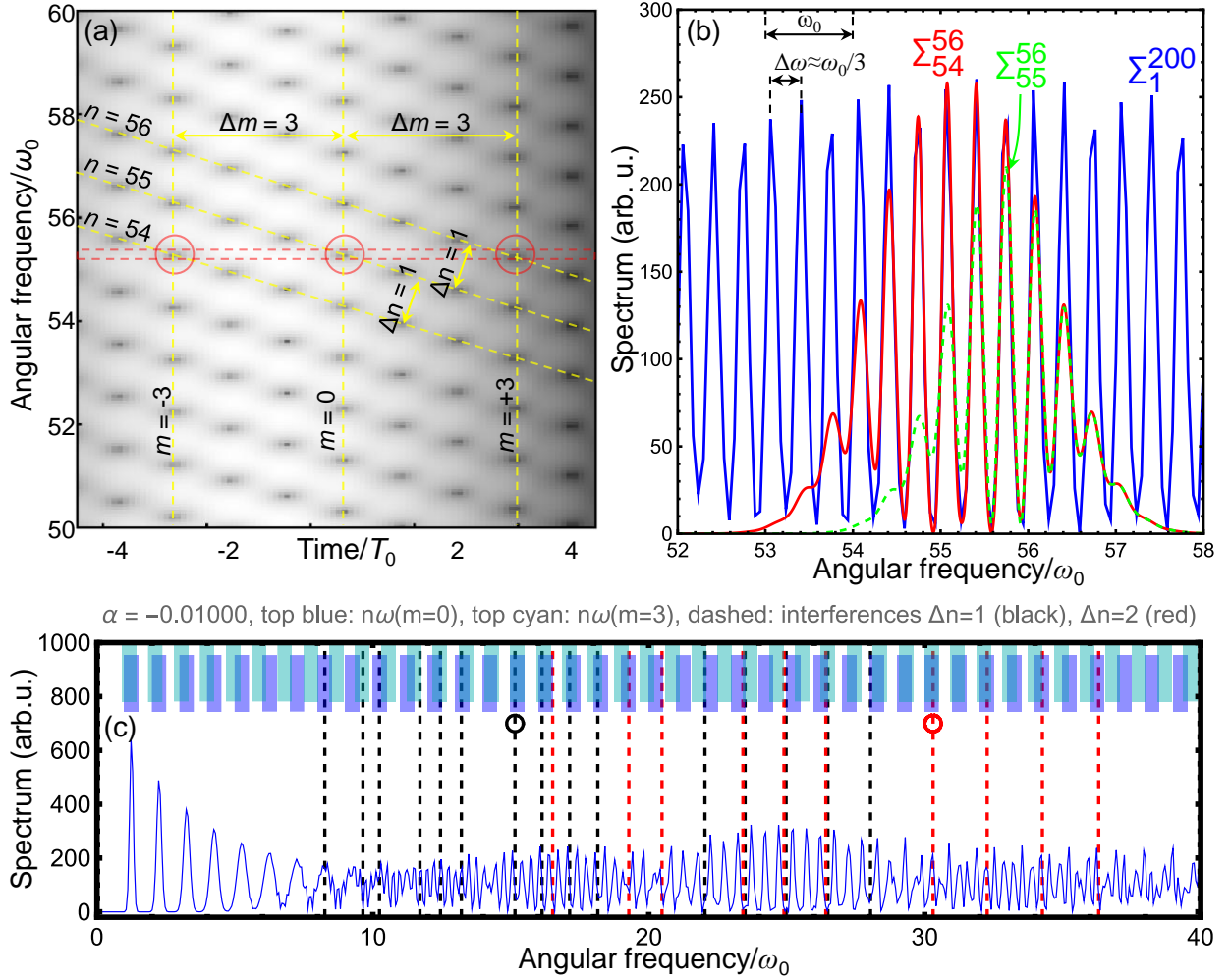

FIG. 7. Alloharmonic generation by a nonperiodic driver according to the model (10) with the linear chirp (15) with the same parameters as in FIG. 2. (a) Portion of a spectrogram. At the region of interest indicated by the red dashed lines at frequency $\omega\sim55\omega_0$, there are three spectrogram peaks indicated by red circles. These spectrogram peaks correspond to three different harmonics with orders $n=54, 55, 56$ (tilted yellow dashed lines) from three different pulses with numbers $m=-3, 0, +3$ (vertical dashed yellow lines). (b) Alloharmonics (interharmonic interference) corresponding to red circles in (a): the spectrum (blue, sum of all harmonics) and its local approximation as a sum of three (red, orders $n=54, 55, 56$) and two (dashed green, orders $n=55, 56$) interfering harmonics with $\Delta m=3$. (c) Frame from the Supplemental Movie [42] showing the variation of the spectrum as $\alpha$ changes from 0 to $-0.01$ (this frame). Blue and cyan bars at the top show the harmonics of the time domain pulses number $m_1=0$ and $m_2=3$, respectively. At frequencies where these bars match, i.e., frequencies providing approximate solutions to the alloharmonic equation (1), the alloharmonics of the $\Delta m=m_2-m_1=3$ series with fringe separations of $\Delta\omega\approx\omega_0/3$ appear (the small black and red circles represent $\Delta n=1$ and $\Delta n=2$, respectively); note that the matches of the blue and cyan bars indicate only positions of these $\Delta m=3$, but not other, alloharmonics. These and other interferences are indicated by vertical dashed lines, with black lines representing interharmonic separation $\Delta n=1$ and red lines representing $\Delta n=2$.

## E. Alloharmonics with a quadratic chirp (cubic phase)

Cubic phase with a quadratic chirp,

$$\Phi(t) = \omega_0 t \left(1 + \beta \left(\frac{t}{T_0}\right)^2\right), \quad (20)$$

$$\omega(t) = \omega_0 \left(1 + 3\beta \left(\frac{t}{T_0}\right)^2\right),$$

arises, e.g., in a frequency comb originating from a cavity with an accelerating mirror. Quadratic chirp produces alloharmonic fringes exhibiting two periods, with separations approximately but not exactly $\Delta\omega \approx \omega_0$ between stronger and $\Delta\omega \approx \omega_0/\Delta m$ between weaker spectral fringes, FIG. 8(a). The former fringes appear due to a strong interference between pulse numbers $m=-1$ and $0$ and $0$ and $+1$ (both $\Delta m=1$), as in this case $t_{-1}=-t_{+1}$, $t_0-t_{-1}=t_{+1}-t_0$, and $\omega_{-1}=\omega_{+1}$, similar to the periodic case (although $t_m$ are $\omega_m$ are different for $|m|>1$).

For the quadratic chirp, the alloharmonic frequencies are $\frac{\omega_A}{\omega_0} \approx -\frac{1}{3\beta}\frac{\Delta n}{\Delta m^2} - \frac{5}{3}\Delta n$, the fringe separations are $\frac{\Delta\omega}{\omega_0} \approx \frac{1}{\Delta m} + \beta\Delta m$, and the required resolving power is $R \approx -\frac{\Delta n}{3\beta\Delta m}$. In contrast to the linear chirp, the required resolving power for the quadratic chirp decreases with $\Delta m$ and thus with the envelope duration or observation time. The frequencies at which alloharmonics appear and become dominant are $\frac{\omega_{\min}}{\omega_0} \approx \frac{1}{12|\beta|}\left(\frac{T_0}{\tau}\right)^2$ and $\frac{\omega^*}{\omega_0} \approx \frac{1}{3|\beta|}\left(\frac{T_0}{\tau}\right)^2$, respectively, FIG. 8(b).

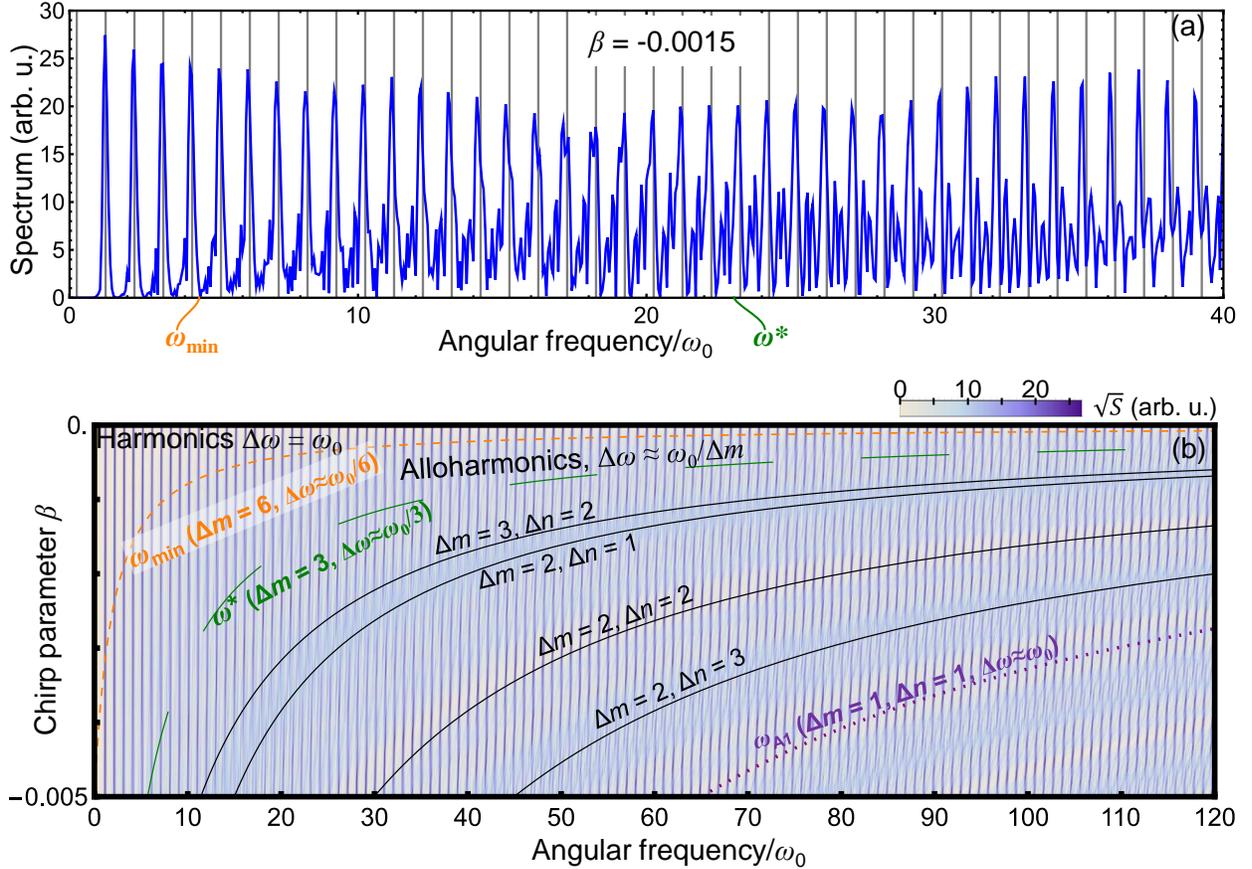

FIG. 8. Alloharmonic generation by a nonperiodic driver according to the nonlinear phase model (10) with the quadratic chirp (20) and other model parameters as in FIG. 2. (a) Spectrum for $\beta=-0.0015$; the grey vertical lines show positions of ideal harmonics, $(n+\delta n)\omega_0$. (b) Variation of the spectrum with $\beta$; the dashed orange line and dashed green line denote the onset ($\omega_{min}$) and dominant ($\omega^*$) region of the alloharmonics, respectively; these frequencies for $\beta=-0.0015$ are also shown in (a). Solid lines show some of the alloharmonics with indicated interharmonic separations $\Delta n$ and pulse separations $\Delta m$ corresponding to fringe separations $\Delta\omega \approx \omega_0/\Delta m$. The purple dotted line shows the $\Delta m=1$ alloharmonics with fringe separation $\Delta\omega \approx \omega_0$.

# APPENDIX C: FITTING OF THE EXPERIMENTAL BISER SPECTRUM WITH THE ALLOHARMONIC MODEL

We used the Simplex and Levenberg Marquardt algorithms in the OriginPro software [84] to fit the experimental BISER spectrum with the alloharmonics.

First, we used the alloharmonic model, Eq.(10) with phase (14) and a Lorentz-like envelope and zero carrier-envelope offset and spectral phase:

$$(21) \quad E(t) = \left(1 + \frac{t^2}{\tau^2}\right)^{-\frac{2}{3}} \sum_n a_n \cos[n\Phi(t)].$$

The summation was performed from the 30$^{th}$ to 250$^{th}$ harmonics. The spectral amplitudes were represented as the double-exponent function $a_n = A_1 \exp(B_1 n\hbar\omega_0) + A_2 \exp(B_2 n\hbar\omega_0)$ to account for the overall spectral shape. The phase contained the chirp coefficients $\alpha$, $\beta$, $\gamma$, and $\delta$ [Eq.(14)], as lower-order phases did not provide good fits to the data. Thus, there were 10 fit parameters: $\tau$, $A_{1,2}$, $B_{1,2}$, $\lambda_0$, $\alpha$, $\beta$, $\gamma$, and $\delta$. Five parameters, $\tau$, $A_{1,2}$, and $B_{1,2}$, fitted the overall spectrum shape and modulation depth and were not alloharmonic-specific. The remaining five parameters, $\lambda_0$, $\alpha$, $\beta$, $\gamma$, and $\delta$, determined the alloharmonic properties. The resulting fitting curve is shown in FIG. 3(a) and FIG. 9(a), the fit parameters and derived parameters in the middle column of Table I.

The dependence of the pulse position shift $\delta t_m = t_m - t_{m,\text{Periodic}} = t_m - mT_0$, on pulse number $m$, had a linear tilt with a slope of $\approx 0.05\omega_0$, FIG. 3(b); here $t_{m,\text{Periodic}} = mT_0$ is the $m^{th}$ pulse position in the periodic case. This might be interpreted as a necessity to change the central frequency by this amount. However, most of the energy was concentrated in the pulses with $m=0,\pm 1$, and thus, the three central pulses were the most important for the central frequency determination. The line passing through the three central points in FIG. 3(b) is approximately horizontal, confirming the correct value of the central frequency inferred from the fit.

To check that the central frequency could not be selected close to the initial laser frequency, we performed a separate fit with the same model, but the central wavelength bounded to [750-1100] nm corresponding to photon energies of [1.13-1.65] eV. It gave a poor fit. We compared this fit with the previous one using the Bayesian information criterion (BIC), which is suitable because the number of data points (844 in our case) was much larger than the number of parameters; the Bayesian information criterion has an advantage as it requires no priors. The comparison showed that the original ($\lambda_0=1964.2\pm 0.3$ nm, $\hbar\omega_0=0.63123\pm 0.00010$ eV) alloharmonic model fit was decisively better, with $\Delta\text{BIC}\approx 10^3$.

We also attempted to fit the experimental data with the normal harmonics, i.e., with a phase with $\alpha=\beta=\gamma=\delta=0$: the result is shown in FIG. 9(b), with the parameters in the right column of Table I. The Bayesian information criterion showed that the harmonic model did not fit the data well, and that the alloharmonic model obtained a decisively better fit, with $\Delta\text{BIC}\approx 6\times 10^2$.

We note that the model spectrum is not sensitive to the overall spectral phase and time direction – i.e., the spectrum does not change when an overall spectral phase is added, or time is reversed. Thus, our fitting procedure did not provide the attosecond pulse duration or the sign of the frequency change. In our case, we selected the time direction so that the frequency decreases gradually because the fundamental frequency of the BISER radiation was ~2.4 times lower than the initial laser frequency, thus requiring a generally decreasing trend.

TABLE I. Experimental spectrum fits with the alloharmonic and harmonic models; error bars of the fit parameters are the fit standard errors [84]; error bars of the derived parameters are calculated by the error propagation [86] taking into account the covariance matrix of the fit parameters.

| Fit parameters | Alloharmonics | Harmonics |
|---|---|---|
| Parameter number | 10 | 6 |
| Reduced $\chi^2$ | 14.9 | 30.9 |
| Adj. R-Square | 0.988 | 0.975 |
| $\lambda_0$ (nm) (central wavelength) | 1964.2±0.3 | 1938.8±0.7 |
| $\tau$ (fs) (envelope duration) | 1.67±0.08 | 0.98±0.10 |
| $A_1$ (arb. u.) (amplitude 1) | 363±30 | 284±31 |
| $B_1$ (slope parameter 1) | −0.105±0.002 | −0.097±0.003 |
| $A_2$ (arb. u.) (amplitude 2) | 0.70±0.11 | 0.38±0.10 |
| $B_2$ (slope parameter 2) | 0.012±0.002 | 0.020±0.004 |
| $\alpha$ (dimensionless linear chirp) | $-(1.684\pm0.007)\times10^{-2}$ | 0 (fixed) |
| $\beta$ (dimensionless quadratic chirp) | $+(3.02\pm0.03)\times10^{-3}$ | 0 (fixed) |
| $\gamma$ (dimensionless cubic chirp) | $+(3.478\pm0.016)\times10^{-3}$ | 0 (fixed) |
| $\delta$ (dimensionless quartic chirp) | $-(3.676\pm0.007)\times10^{-3}$ | 0 (fixed) |
| Derived parameters | | |
| $\hbar\omega_0$ (eV) (central photon energy) | 0.63123±0.00010 | 0.6395±0.0002 |
| $T_0$ (fs) (central cycle duration) | 6.5517±0.0011 | 6.467±0.002 |
| $t_{-2}/T_0$ (position of the $m=-2$ pulse) | −2.1253±0.0002 | −2 (fixed) |
| $t_{-1}/T_0$ (position of the $m=-1$ pulse) | −0.98743±0.00006 | −1 (fixed) |
| $t_{+1}/T_0$ (position of the $m=+1$ pulse) | +1.01444±0.00006 | +1 (fixed) |
| $t_{+2}/T_0$ (position of the $m=+2$ pulse) | +2.1394±0.0002 | +2 (fixed) |
| $b_{-2}$ (amplitude of the $m=-2$ pulse) | 0.032±0.002 | 0.018±0.002 |
| $b_{-1}$ (amplitude of the $m=-1$ pulse) | 0.088±0.006 | 0.044±0.006 |
| $b_{+1}$ (amplitude of the $m=+1$ pulse) | 0.085±0.005 | 0.044±0.006 |
| $b_{+2}$ (amplitude of the $m=+2$ pulse) | 0.032±0.002 | 0.018±0.002 |

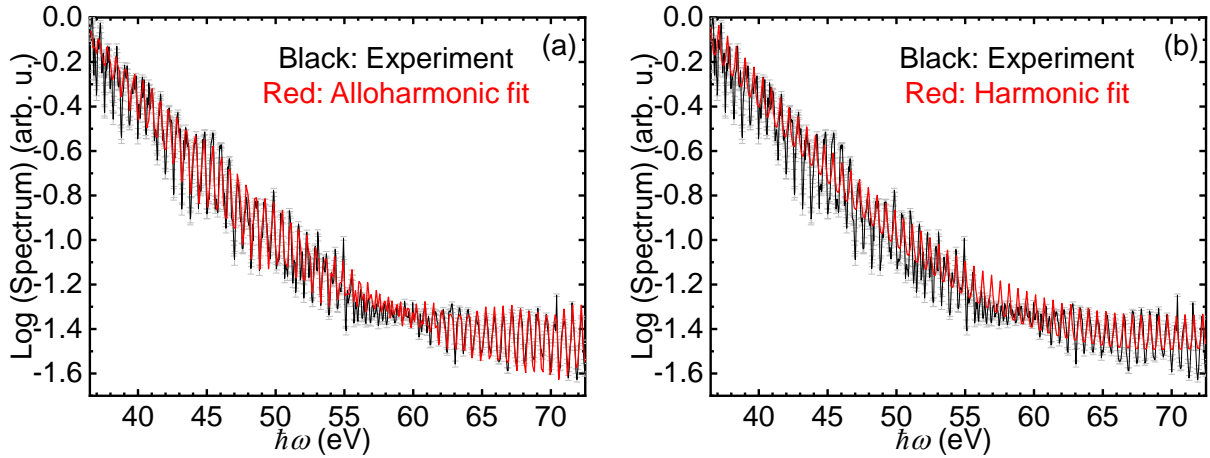

FIG. 9. Fits (red) of the experimental BISER spectrum (black) with (a) alloharmonic model and (b) harmonic model; the black curve is the same in (a), (b), and FIG. 3(a). Error bars in the experimental spectrum are estimated standard deviations including the spectrograph's CCD dark current, readout, and shot noise. The fit parameters are shown in Table I.

# APPENDIX D: NUMERICAL RELATIVITY CALCULATION OF GRAVITATIONAL WAVES EMITTED BY BINARY BLACK HOLE MERGERS

We used gravitational waveforms obtained by numerical relativity calculations from the Simulating eXtreme Spacetimes (SXS) Collaboration catalogue [43, 44]. Specifically, we used center-of-mass-corrected, non-extrapolated spherical harmonic modes from the outermost extraction radius. We calculated the strain from these modes using expressions for the spin-weighted spherical harmonics from [87]. We selected the in-plane observation direction because it contains much stronger harmonics than other observation directions. Details are in the Supplementary Material [45].

The harmonic and alloharmonic spectra shown in FIG. 4(c,d) were calculated as the Fourier transforms of the strain in time windows shown in FIG. 4(a) by magenta and black lines, respectively. To compare these numerically calculated spectra with the analytical alloharmonic model, we calculated the time-dependent orbital frequency, as shown by the blue line in FIG. 4(a), using interpolated zero-crossings of the strain and calculating the orbital period as four times the separation between these zero-crossings. The coefficient of 4 is necessary because there are two crossings per period, and the main gravitational wave contribution is emitted at the second harmonic of the orbital frequency. We checked that another method based on the separations between the positive and negative peaks of the strain produced consistent results [45].

The dependence of the orbital frequency on time was used to determine the local central frequency $\nu_0 = \omega_{Orb}/2\pi$ and chirp coefficients $\alpha$, $\beta$ for the slowly varying frequency in the magenta time window in FIG. 4(a), and $\alpha$, $\beta$, $\gamma$ for the faster-varying frequency in the black time window in FIG. 4(a). These local central frequencies were used to calculate the frequencies of the normal harmonics $n\nu_0$ shown by the solid vertical lines in FIG. 4(c,d), for which the 2$^{nd}$ harmonic dominates. The chirp coefficients were used to calculate the frequencies of the alloharmonic series using Eq.(1), shown by dashed vertical lines in FIG. 4(d).

In the Supplemental Movie [46], the orbital frequency gradually increases with time. Thus, the number of gravitational wave cycles in the constant-duration time window also increases. This differs from the Supplemental Movie [42], where the number of wave cycles is constant. Therefore, in [46], when the alloharmonics first appear in the shown spectral band, they have a maximum $\Delta m = 4$, while as the inspiral progressed, this value increases up to $\Delta m = 8$. We set the animation frame rate to 8 frames per second (FPS), i.e., the animation frame duration of 125 ms. The physical frame duration in the animation is $100M$ (geometric units), thus, the 125 ms frame duration corresponds to $M \approx 254 M_\odot$ (For $M = M_\odot$, the geometric time unit is 4.93 μs). For a combined binary mass of $M \approx 25 M_\odot$, the animation would be ten times faster.

# APPENDIX E: ALLOHARMONICS IN RADIO SPECTRA OF PULSARS

Table II. Predicted parameters of alloharmonics in radio spectra of pulsars.

| Pulsar | Age, years | $\nu_0$, Hz | $\dot{\nu}$, Hz$^2$ | $\alpha$ | $\tau$, s | $N_p$ | $n_{min}$ | $\nu_{min}$, MHz | $R_{min}$ | $\alpha_E$ | $\alpha_{L2}$ |
|---|---|---|---|---|---|---|---|---|---|---|---|
| J0835-4510 Vela | $1.13 \times 10^4$ | 11.1946 | $-1.5666 \times 10^{-11}$ | $-6.25 \times 10^{-14}$ | 205 | 2300 | $3.48 \times 10^9$ | $38.9 \times 10^3$ | $8.00 \times 10^{12}$ | $-5.05 \times 10^{-12}$ | $-8.94 \times 10^{-13}$ |
| Vela, 2016 glitch | | | | $+2.32 \times 10^{-9}$ | 4.4 | 49 | $4.39 \times 10^6$ | 49 | $2.15 \times 10^8$ | | |
| J0633+1746 Geminga | $3.42 \times 10^5$ | 4.2176 | $-1.9516 \times 10^{-13}$ | $-5.49 \times 10^{-15}$ | 205 | 864 | $1.05 \times 10^{11}$ | $445 \times 10^3$ | $9.12 \times 10^{13}$ | $-1.34 \times 10^{-11}$ | $-2.37 \times 10^{-12}$ |
| J1808-2024 SGR 1806-20 | 218 | 0.13234 | $-9.62 \times 10^{-12}$ | $-2.75 \times 10^{-10}$ | 204 | 27 | $6.74 \times 10^7$ | 8.92 | $1.82 \times 10^9$ | $-4.27 \times 10^{-10}$ | $-7.56 \times 10^{-11}$ |
| Same, shorter time | -//- | -//- | -//- | -//- | 15 | 2 | $9.10 \times 10^8$ | 120 | -//- | -//- | -//- |
| J162759.5-523504.3 | | 0.000916 | $-5.04 \times 10^{-16}$ | $-3.00 \times 10^{-10}$ | 3274 | 3 | $5.56 \times 10^8$ | 0.509 | $1.67 \times 10^9$ | $-6.17 \times 10^{-8}$ | $-1.09 \times 10^{-8}$ |

$v_0$ and $\dot{v}$: Pulsar rotation frequency and its time derivative (data from Ref.[55] for all rows except the bottom row and Ref.[88] for the bottom row), α: dimensionless linear chirp parameter, τ and $N_P$: observation time window and number of periods in this window, $n_{min}$ and $v_{min}$: minimum harmonic order and frequency above which alloharmonics appear, $R_{min}$: required resolving power to observe alloharmonics, $α_E$ and $α_{L2}$: largest possible dimensionless chirp parameters of alloharmonics caused by the Earth's rotation and orbit at the L2 Sun-Earth Lagrange point [54].